\documentclass[aps,amsmath,amssymb,showpacs,twocolumn,longbibliography, pra]{revtex4-1}
\usepackage{hyperref}
\usepackage{epsfig}
\usepackage{subfigure}
\usepackage{braket} 
\usepackage{color}
\usepackage{verbatim}
\usepackage{braket}

\newcommand{\pardiff}[2]{\frac{\partial{#1}}{\partial{#2}}}
\newcommand{\ad}{\hat{a}^{\dagger}}
\newcommand{\bd}{\hat{b}^{\dagger}}

\newcommand{\Gu}{\Gamma_{\uparrow}}
\newcommand{\Gd}{\Gamma_{\downarrow}}
\newcommand{\tGu}{{\Gamma}_{\uparrow}^{\rm tot}}
\newcommand{\tGd}{{\Gamma}_{\downarrow}^{\rm tot}}
\newcommand{\tGun}{{\Gamma}_{\uparrow}^{\rm tot}(n_m)}
\newcommand{\tGdn}{{\Gamma}_{\downarrow}^{\rm tot}(n_m)}

\newcommand{\omDye}{\omega_{D}}

\newcommand{\modelabel}{m}
\newcommand{\nl}{n_\modelabel}
\newcommand{\average}[1]{\langle{#1}\rangle}
\renewcommand{\vector}[1]{{\boldsymbol{#1}}}

\newcommand{\maxnm}{\max\{n_m\}}
\newcommand{\ommax}{\omega_{m_{\rm max}}}

\newcommand{\ua}{\uparrow}
\newcommand{\da}{\downarrow}

\usepackage[normalem]{ulem} 

\begin{document}

\title{Thermalization and breakdown of thermalization in photon condensates}

\author{Peter Kirton}
\affiliation{SUPA, School of Physics and Astronomy, University of St Andrews, St Andrews, KY16 9SS, United Kingdom}
\author{Jonathan Keeling}
\affiliation{SUPA, School of Physics and Astronomy, University of St Andrews, St Andrews, KY16 9SS, United Kingdom}
\date{\today}

\begin{abstract}
  We examine in detail the mechanisms behind thermalization and
  Bose-Einstein condensation of a gas of photons in a dye-filled
  microcavity. We derive a microscopic quantum model, based on
    that of a standard laser, and show how this model can reproduce
    the behavior of recent experiments.  Using the rate equation
    approximation of this model, we show how a thermal distribution of
    photons arises.  We go on to describe how the non-equilibrium effects
    in our model can cause thermalization to break down as one moves
    away from the experimental parameter values.  In particular, we
    examine the effects of changing cavity length, and of altering the
    vibrational spectrum of the dye molecules.  We are able to
    identify two measures which quantify whether the system is in 
    thermal equilibrium.  Using these we plot ``phase diagrams''
    distinguishing BEC and standard lasing regimes.  Going beyond the
    rate equation approximation, our quantum model allows us to
    investigate both the second order coherence, $g^{(2)}$, and the linewidth of the emission from the cavity. We show
    how the linewidth collapses as the system transitions to a Bose condensed
    state, and compare the results to the Schawlow--Townes linewidth.
\end{abstract}

\maketitle

\section{Introduction}

Recent experiments have convincingly demonstrated the Bose-Einstein
condensation~\cite{pitaevskii03} of gases of photons, both as dressed
photons (exciton-polaritons)~\cite{Kasprzak2006,Balili2007}, and,
more recently, of pure photons in a dye-filled
microcavity~\cite{Klaers2010c}.  Such quantum fluids of
light~\cite{Carusotto2012} reinvigorate investigation of the
relation between condensation and lasing~\cite{Graham1970a,haken75}.  In a
dye-filled microcavity, photons can establish a thermal distribution
by repeated absorption and re-emission~\cite{Klaers2010b}.  However,
in these systems the steady state is not purely defined in terms of
the energetics; the unavoidable losses and pumping mean that the
non-equilibrium nature of the experiments must be taken into account.
As an open system emitting coherent light, there is an evident
connection to a laser, but the observation of a Bose-Einstein
distribution clearly suggests that more is going on than standard
lasing.  The aim of this work is to present in detail a quantum
mechanical model which addresses exactly this question: when does a
dye-filled cavity behave as a standard laser, and when does it behave
as a condensate?

 The paradigmatic examples of a textbook laser~\cite{haken70} and a textbook Bose-Einstein
  condensate~\cite{pitaevskii03} are quite distinct: in the textbook
  laser, the population of modes is controlled by gain and loss, and
  lasing occurs when linear gain exceeds the loss rate.  In the
  textbook BEC, the population of modes is controlled by their
  energies, according to the Bose-Einstein distribution, and
  condensation occurs when the chemical potential reaches the lowest
  mode. However, this distinction is less clear cut than may first
  appear: the quantum Boltzmann equation describes the rates of
  scattering into and out of a given mode, and its steady state
  describes a Bose-Einstein distribution~\cite{Lifshitz1981}.  Thus,
  there can be situations where, when the scattering rates depend on
  energy, one recovers the equilibrium thermal
  distribution~\cite{doan05:prb,Doan2006}. 
  
  In its traditional setting lasing is considered for a single- or
  few-mode cavity, while Bose-Einstein condensation is considered in a
  spatially extended system.  This distinction is however absent for
  wide aperture lasing systems, such as vertical cavity surface
  emitting lasers (VCSEL).  Experiments on photon and polariton
  condensates also typically involve structures with multiple
  transverse photon modes.  As such, in these extended systems the
  question of the transition between lasing and condensation has been
  of considerable interest, leading to extensive
  discussion~\cite{Butov2007,Snoke2008,Fischer2013}, as well as many
  theoretical and experimental works exploring this crossover for
  polaritons~\cite{deng03,Malpuech2003,doan05:prb,Szymanska2006,Doan2006,Szymanska2007,Dang2008,Keeling2013,Haug2014}.
  For photons, the question has been less extensively studied,
  preliminary results were given in our earlier
  Letter~\cite{Kirton2013b}, in the current work we use the same model to
  discuss the lasing-condensation crossover in more detail, and
  explore quantum effects beyond the rate equation approximation.
  
  As well as the archetypal examples of a textbook laser or
  Bose-Einstein condensate, there are several other recent examples of
  photonic systems which show phase transitions that can be related to
  condensation.  While, as we will discuss below, these are quite
  distinct from the behavior seen in the dye-filled microcavity, it is
  illuminating to understand what these differences are, and to
  place experiments in dye-filled microcavities within the wider
  landscape of condensation of light.

  One example of condensation of light concerns the statistical
  description of mode locking in
  lasers~\cite{Weill2010,Rosen2010,Schwartz2013,Fischer2012,Fischer2013}.
  This system is notable in that it does not require multiple
  transverse modes, but rather concerns different temporal modes in
  pulsed lasing.  In particular, if active mode locking (AML) is
  described in terms of the time-dependent eigenmodes, $\psi_m$, of
  the modulation profile, the occupations of each eigenmode obey a
  linearized Langevin equation, $\partial_t \psi_m = (\eta-\kappa_m)
  \psi_m + \Gamma_m$, where $\kappa_m$ is a decay rate of a given
  mode, $\eta$ is an overall linear gain, and $\Gamma_m$ a noise term.
  If $\eta$ and $\kappa_m$ are regarded as freely adjustable, this
  model would show instability whenever $\eta > \kappa_m$; physically,
  however, gain saturation means that the effective gain, $\eta$,
  decreases as the mode population increases.  If one views this gain
  saturation as adjusting the parameter $\eta$ such that the total
  power, $\sum_m |\psi_m|^2$ is fixed then this equation can show
  condensation~\cite{Weill2010}, i.e.\ there can be a transition to a
  state where the mode with smallest $\kappa_m$ acquires a macroscopic
  occupation.  Whether or not a transition occurs is controlled by the
  density of states of eigenmodes, as expected for condensation.
  However, in this system the relevant density of states is the
    density per interval of decay rates, $g(\kappa)$, so that the
    number of modes having decay rates in the range $[\kappa, \kappa
    + d\kappa]$ is given by $g(\kappa) d\kappa$.  This means that
  condensation occurs if there are relatively few long lived modes,
  but not if the density of long lived modes is too high. The density
  of states can be varied by changing the modulation
  profile~\cite{Weill2010}, as has been experimentally
  observed~\cite{Rosen2010}.  There also exist methods to vary the
  density of states by modulating with a ``hypercomb'', involving
  multiple incommensurate frequency components, changing the
  connectivity (dimensionality) of the mode space~\cite{Schwartz2013}.
  These ideas of how mode locking can be understood as condensation
  are reviewed in Refs.~\cite{Fischer2012,Fischer2013}.

  The AML phase transition described in~\cite{Weill2010,Rosen2010} can
  be viewed as condensation, but as well as the oddity that it is a
  density of states in loss rate, not energy, which controls the
  distribution, a second notable difference appears compared to the textbook BEC.
  This is the fact that the distribution takes the form $n_m \propto
  T/(\kappa_m - \eta)$, with $\kappa_m$ the loss rate, and $\eta$ the
  gain, which plays the role of chemical potential.  This matches the
  form of the low energy expansion of a Bose distribution $n_m =
    n_B(\epsilon_m) = [\exp((\epsilon_m-\mu)/T) -1]^{-1} \simeq T /
    (\epsilon_m-\mu)$, however the AML distribution is not simply a low
  frequency approximation, but rather describes the actual distribution.  As
  such, the distribution is not Bose-Einstein, but rather
  Rayleigh-Jeans, and the AML condensate is thus best described as a
  Rayleigh-Jeans condensate.  Intriguingly, this effect has been
  studied in other contexts, both theoretically and experimentally.
  Theoretically, such an object arises in classical field
  methods~\cite{Blakie:Dynamics,Griffin2009a} where a finite lattice
  resolution is used to cut off high momentum states. An experimental verification of this
  comes from the experiments of \citet{Sun2012} which showed
  ``condensation of classical light'', again a Rayleigh-Jeans
  condensate, by passing light through a strongly non-linear medium.

  In contrast to the Rayleigh-Jeans condensates, the experiments on
  dye-filled microcavities show not only condensation, but also a
  Boltzmann tail and so the condensation is related to the
  Bose-Einstein distribution as distinct from the Rayleigh-Jeans.  The
  essential ingredients required for this to occur in the dye
  filled microcavity are absorption and re-emission of photons by the
  dye molecules.  As such, the condensate is formed by stimulated
  emission of radiation, yet, as we will discuss below, this can be
  associated with a Bose-Einstein distribution, including the high
  energy Boltzmann tail.  The mechanism leading to this thermal
  distribution is quite distinct from that in cold atoms or
  polaritons, where direct atom-atom or polariton-polariton
  interactions exist.  Nonetheless, as we will discuss, for small
  enough cavity loss rates, the process of repeated absorption and
  re-emission of photons can establish a thermal distribution.  As
  such, despite the differing mechanism, the observable properties of
  the dye-filled cavity can be identical to that of an equilibrium
  BEC.  To understand the distinctions it is therefore of particular
  interest to understand the behavior as thermalization breaks down,
  as discussed in this paper.

  Since the initial observation of condensation in dye-filled
  microcavities, further experimental work has probed thermalization
  of light in other media~\cite{Klaers2011a,Schmitt2012a}, the
    statistics of condensate fluctuations~\cite{Schmitt2014}, the
    role played by the size of the pumping spot~\cite{Marelic2014} and the possibility of a lasing to condensation crossover~\cite{Schmitt2014b}.
  Inspired by these experiments, there has also been significant
  theoretical work on a variety of topics related to photon
  condensation.  Many of these works have concentrated on the unique
  properties of the photon system even in thermal
  equilibrium~\cite{Klaers2012a, Sob'yanin2012,
    Kruchkov2014,VanderWurff2014}. These have included exploration of
  the role of the dye molecules as a reservoir for excitations,
  leading to grand canonical
  statistics~\cite{Klaers2012a,Kruchkov2014}, and exploring effects of
  the nonlinearity of coupling to dye molecules inducing effective
  interactions~\cite{VanderWurff2014}.  Other work has studied the
  dynamics resulting from photon-phonon scattering, and how this may
  lead to a Bose-Einstein distribution~\cite{Snoke2012} in the absence
  of loss.  More recently, aspects of photon condensation including
  loss have been considered, including a derivation of an effective
  dissipative order parameter equation from interactions induced by
  the dye molecules~\cite{Nyman2014}. The phase correlations,
  including effects of photon loss and interactions on the time and
  space correlations of the condensate phase, have also been
  explored~\cite{DeLeeuw2014,Chiocchetta2014,DeLeeuw2014a}.  Several
  of these questions have been very similarly addressed in the
  literature for polariton or atom lasers.  For example, phase
  diffusion due to interactions was studied
  by~\cite{tassone00,porras03}, and the effect of particle loss on
  phase correlations in a dissipative condensate has been extensively
  studied~\cite{Szymanska2006,wouters06,Szymanska2007,Roumpos2012,Altman2013b,Chiocchetta2014,Gladilin2014}.
  However, none of these works have started from the microscopic model
  of a dye-filled microcavity accounting for the vibrational modes of
  the dye molecules, and thus do not fully describe the mechanisms
  that apply in the
  experiments~\cite{Klaers2010b,Klaers2010c,Schmitt2014}.

  In this paper we develop further a microscopic model for the photon
  condensate system, as introduced in our previous
  work~\cite{Kirton2013b}: we consider a series of photon modes
  coupled to electronic excitations of dye molecules which are in turn
  coupled to a ladder of rovibrational states. These provide the
  thermal equilibrium bath necessary to observe BEC. We examine in
  detail the mechanisms behind the thermalization processes, and show
  how this leads to the formation of a BEC inside the cavity. We also
  show how, by changing the parameters of either the cavity or the
  dye, this mechanism can break down and lead to non-thermal
  behavior.  Going beyond the rate equation treatment we previously
  presented~\cite{Kirton2013b}, we also discuss features requiring the
  full quantum model, such as the linewidth of the photon condensate 
and the second order coherence.

The structure of this paper is as follows. In Sec.~\ref{sec:model} we
present in detail the derivation of the quantum mechanical model
  which we introduced in our Letter, Ref.~\cite{Kirton2013b}.  We
  also discuss extending this model to include multiple rovibrational
  modes of the molecules.  After this introduction, we then divide
  discussion of the results into two sections. Section~\ref{sec:rate}
  discusses those features of the experiments which can be understood
  within a rate equation model, derived from the full quantum
  description of the system.  After deriving the rate equation, and
  discussing the condensation threshold condition in
  Sec.~\ref{sec:deriv-rate-equat}, we go on, in Sec.~\ref{sec:breakdown},
  to use the rate equation to discuss the ways in which the
  thermalization process can break down.  We consider the effect of
  changing the cavity cutoff frequency, the thermalization rate of the
  dye, the coupling between vibrational and electronic states, and the
  temperature of the system.  In Sec.~\ref{sec:dynam-therm} we apply
  the rate equation to consider the dynamics of thermalization after
  an initial excitation.  Section~\ref{sec:linewidths} returns to the
  full quantum model, and discusses how we can go beyond the rate equation approach and use this to calculate both the second order coherence, $g^{(2)}$, and
  the linewidth of the condensed mode as it passes through the
  threshold.  Finally, in Sec.\ \ref{sec:conc}, we present our
  conclusions.

\section{Model} \label{sec:model}

\begin{figure}
  \centering
  \includegraphics[width=2.5 in]{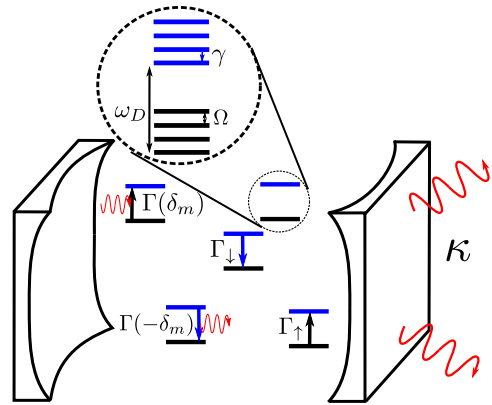}
  \caption{(Color online) Cartoon of the system showing the decay
    processes included in Eq.~\eqref{eqn:ME}. The zoomed in view shows
    the energy level structure of the dye molecules. 
    }\label{fig1}
\end{figure}

A schematic diagram of the system we consider is shown in
Fig.~\ref{fig1}. It consists of a set of cavity modes coupled
to a solution of dye molecules, modeled as described in detail below.

The cavity used in the experiments confines the photons in a two
dimensional plane and imposes a harmonic trap on the condensate. This
allows us to specify a set of evenly spaced modes, $\omega_m =
\omega_0 + m \epsilon$, with the lowest energy mode at $\omega_0$ and
spacing $\epsilon$. In two dimensions, the mode with index $m$ has
degeneracy $g_m=m+1$. A photon in this mode is created by the operator
$\ad_m$. If this photon gas is in thermal equilibrium, the low energy cut-off along with the two-dimensional
harmonic oscillator level spacing and degeneracies provide the
necessary conditions to observe a BEC~\cite{pitaevskii03}.

Each dye molecule, labeled by the index $i$, is modeled as a two-level
system (corresponding to its electronic state), and a bosonic mode (or
modes) describing its vibrational state.  Operators on the electronic
states are written in terms of Pauli matrices $\hat{\sigma}_i$ and we
denote the bare splitting between ground and excited electronic states
(i.e.\ without vibrational dressing) as $\omDye$.  Each electronic
state is broadened into a ladder of rovibrational states.  In the
simplest case these correspond to the modes of an harmonic oscillator
with frequency $\Omega$.  As discussed below, more complicated
configurations can be described by incorporating multiple
rovibrational modes. We denote the creation operators for this
harmonic oscillator mode as $\bd$.  The coupling constant between
electronic and rovibrational degrees of freedom is parameterised by
the Huang-Rhys factor $S$. This corresponds to the relative oscillator
displacement between the ground and excited manifolds in units of the
harmonic oscillator length of the given mode.

The photon modes are coupled to the electronic transition by means of a standard Jaynes-Cummings interaction with coupling constant $g$ which we assume to be weak throughout. Combining all of this, the Hamiltonian is thus
\begin{multline}
  \label{eqn:H}
  \hat{H}=\sum_{\modelabel}\omega_\modelabel\ad_\modelabel \hat{a}_\modelabel +
  g\sum_{\modelabel,i}\left(\hat{a}_\modelabel\hat{\sigma}_i^++\ad_\modelabel\hat{\sigma}_i^-\right)
  \\
  +\sum_i\frac{\omDye}{2}\hat{\sigma}_i^z
  +\Omega\left(\bd_i\hat{b}_i+\sqrt{S}\sigma_i^z(\hat{b}_i+\bd_i)\right),
\end{multline}
using units such that $\hbar=1$.

\subsection{Eliminating vibrational modes}

If the coupling to rovibrational modes, $S$, is reasonably strong, then multiphonon
effects will be important in describing the thermalization processes.
To capture these effects it is convenient to make a polaron
transformation $\hat{H}\to \hat{U}^\dagger \hat{H} \hat{U}$, where
\begin{equation}
	\hat{U}=\exp\left[\sum_i\sqrt{S}\hat{\sigma}_i^z(\hat{b}_i-\bd_i)\right].
\end{equation}
This results in a Hamiltonian of the form (ignoring unimportant constants),
\begin{multline}\label{eqn:Hpolaron}
 H=\sum_{\modelabel}\omega_\modelabel\ad_\modelabel \hat{a}_\modelabel+\sum_i\frac{\omDye}{2}\hat{\sigma}_i^z+\Omega\bd_i\hat{b}_i\\
 +{g}\left(\hat{a}\hat{\sigma}_i^+ \hat{D}_i+\ad\hat{\sigma}_i^-\hat{D}_i^\dagger\right),
\end{multline} 
where the displacement operator at site $i$ is
$\hat{D}_i=\exp[2\sqrt{S}(\bd_i-\hat{b}_i)]$.  Since the coupling of
molecules to the optical modes is weak, we then treat the dynamics
perturbatively in $g$ while keeping all orders of $S$. To do this we
write the Liouville equation for the reduced density operator treating
the on site vibrational mode as a bath. Making the standard
Born-Markov approximations as well as secularizing the resulting
equation by removing any terms which oscillate quickly in the
interaction picture one then arrives at the master
equation~\cite{Marthaler2011},
\begin{multline}
  \label{eqn:MEK}
  \dot{\hat{\rho}} =-i[\hat{H}_0,\hat\rho] -\sum_{i,\modelabel}\left\{
    \frac{\kappa}{2}\mathcal{L}[\hat{a}_\modelabel]+\frac{\Gu}{2}\mathcal{L}[\hat{\sigma}_i^+]+\frac{\Gd}{2}\mathcal{L}[\hat{\sigma}_i^-]\right\}\hat\rho\\
   +K(-\delta_\modelabel)\left[\hat{a}_\modelabel\hat{\sigma}^+_i, \ad_\modelabel\hat{\sigma}^-_i\hat{\rho}\right]+ K^{\ast}(-\delta_\modelabel)\left[\hat\rho \hat{a}_\modelabel\hat{\sigma}^+_i, \ad_\modelabel\hat{\sigma}^-_i\right]\\
    +K(\delta_\modelabel)\left[\ad_\modelabel\hat{\sigma}^-_i, \hat{a}_\modelabel\hat{\sigma}^+_i\hat{\rho}\right]+ K^{\ast}(\delta_\modelabel)\left[\hat\rho \ad_\modelabel\hat{\sigma}^-_i, \hat{a}_\modelabel\hat{\sigma}^+_i\right].
\end{multline}
The function $K(\delta)$ will be defined below, and $\delta_m$
  represents the detuning between a given cavity mode and the bare dye
  frequency, $\delta_m=\omega_m-\omDye$.  In writing
  Eq.~(\ref{eqn:MEK}) we have also included additional Markovian loss
terms which describe leakage from the cavity at rate $\kappa$ (assumed
to be identical for all photon modes), incoherent pumping of molecules
to the excited electronic state at rate $\Gu$ and incoherent decay to
the ground state at rate $\Gd$ which describes fluorescence processes
which emit photons into non-cavity modes. These are described by the
usual Lindblad superoperator defined as
$\mathcal{L}[\hat{X}]\hat\rho=\{\hat{X}^{\dagger}
\hat{X},\hat\rho\}-2\hat{X}\hat\rho \hat{X}^{\dagger}$.

The function $K(\delta)$ which appears in the vibration induced terms
is given by the Fourier transform of the retarded correlation
function of displacement operators, broadened by (convolved with)
the incoherent pumping and decay of the electronic degrees of
freedom~\cite{Marthaler2011}:
\begin{equation} \label{eqn:Gdelta}
	K(\delta)=g^2\int_{0}^\infty dt \average{\hat{D}^\dagger_i(t)\hat{D}^{}_i(0)}{\rm e}^{-(\Gu+\Gd)|t|/2}{\rm e}^{-i\delta t}.
\end{equation}
Here $\hat{D}_i(t)$ is the displacement operator in the interaction
picture.  The real parts of this function can be collected into
Lindblad terms which give rise to decay processes which simultaneously
(de)excite a molecule and (emit) absorb a photon. These processes,
along with the other gain and loss terms, are shown schematically in
Fig.~\ref{fig1}. The imaginary parts (Lamb shifts) on the other
hand can be absorbed into the Hamiltonian evolution to give the master
equation,
\begin{multline}
  \label{eqn:ME}
  \dot{\hat\rho} =-i[\tilde {\hat{H}}_0,\hat\rho] -\sum_{i,\modelabel}\left\{
    \frac{\kappa}{2}\mathcal{L}[\hat{a}_\modelabel]+\frac{\Gu}{2}\mathcal{L}[\hat{\sigma}_i^+]+\frac{\Gd}{2}\mathcal{L}[\hat{\sigma}_i^-]\right.\\
    \left.+\frac{\Gamma(-\delta_\modelabel)}{2}\mathcal{L}[\ad_\modelabel\hat{\sigma}^-_i]+\frac{\Gamma(\delta_\modelabel)}{2}\mathcal{L}[\hat{a}_\modelabel\hat{\sigma}^+_i]\right\}\hat\rho.
\end{multline}
 The rate $\Gamma(\delta)$ arises from the real part of the correlation function of displacement operators $\Gamma(\delta)=2{\rm Re}[K(\delta)]$. The Hamiltonian in this equation has been renormalized by the interaction with the bath provided by the vibrational degrees of freedom
\begin{equation}
  \label{eq:lamb_hamil}
	\tilde {\hat{H}}_0 = \sum_{\modelabel,i} \tilde\delta_\modelabel
\ad_\modelabel \hat{a}_\modelabel+ \eta_\modelabel\ad_\modelabel \hat{a}_\modelabel\hat{\sigma}_i^+\hat{\sigma}_i^-.
\end{equation}
In this expression we have shifted into a frame rotating at the
frequency of the bare molecular transition so that only the
  detunings $\delta_m$, and not the individual values of $\omega_m$,
  $\omDye$ appear.  The energy shifts in the Hamiltonian above are
given by $\eta_\modelabel={\rm
  Im}[K(-\delta_\modelabel)-K(\delta_\modelabel)]$ and $\tilde
\delta_\modelabel =\delta_\modelabel+{\rm Im}[K(\delta_\modelabel)]$.
Since, in the photon number basis, Eq.~\eqref{eq:lamb_hamil} is
diagonal (it only couples populations to other populations), these
Lamb shifts do not affect the dynamics at order $g^2$.  As our
approximation is based on expanding in powers of the small parameter
$g/\omega_0$, these Lamb shifts can therefore be ignored, at least
below or at threshold.

The displacement operator correlation function which is required to
find the decay rates in the master equation can be calculated exactly
by considering the Schwinger-Keldysh path integral:
\begin{equation}
  \average{\hat{D}^\dagger_i(t)\hat{D}^{}_i(0)}=\int\mathcal{D}(b\bar{b}){\rm e}^{i\tilde S}
  D^{\ast}_i(t)D^{}_i(0),
\end{equation}
where the action, $\tilde{S}$, is given by
\begin{equation}
  \tilde{S}=\int \frac{d\nu}{2\pi}\;\vector{\bar{b}}\,G^{-1}\,\vector{b}.
\end{equation}
This is written in terms of the inverse Green's function, $G^{-1}$,
  for an harmonic oscillator coupled to a thermal bath.  Writing the
  fields in the Keldysh rotated basis, $\vector{b}=(b_{cl}, b_{q})^T$,
  the inverse Green's function takes the form~\cite{kamanev}:
\begin{equation} \label{eqn:inverseGF}
	G^{-1}=\begin{pmatrix}
	       	0 & \nu-\Omega-\frac{i\gamma}{2} \\ \nu-\Omega+\frac{i\gamma}{2} & i\gamma \coth\left(\frac{\beta\nu}{2}\right) 
	       \end{pmatrix},
\end{equation}
where $\beta=1/k_B T$ is the inverse temperature.
To proceed we write the correlation function in the form:
\begin{equation}
  \average{\hat{D}^\dagger_i(t)\hat{D}^{}_i(0)}=
  \int\mathcal{D}(b_i,\bar{b}_i){\rm e}^{i\tilde S+Q},
\end{equation}
where $Q$ corresponds to the sum of the exponents from
  $D_i^\ast(t)$ and $D_i^{}(0)$.  After completing the square and
carrying out the Gaussian integral this gives
\begin{multline} \label{eqn:ftgen}
  \average{\hat{D}^\dagger_i(t)\hat{D}^{}_i(0)}=\\\exp\left[-i\int\frac{d\nu}{2\pi}|q(\nu)|^2\left(G^A+G^R+G^K\right)\right],
\end{multline}
where $q(\nu)=\sqrt{2S}[\exp(i\nu t)-1]$ and $G^{A/R/K}$ are the advanced, retarded and Keldysh Green's functions respectively.
From the inverse of Eq.~\eqref{eqn:inverseGF} we find that the correlation function is~\cite{Wilson-Rae2002b,
  McCutcheon2011b, Marthaler2011}
\begin{multline} \label{eqn:DtD0}
	\average{\hat{D}^\dagger_i(t)\hat{D}^{}_i(0)}=\\\exp\left[-\frac{2S\gamma}{\pi}\int_{-\infty}^\infty d\nu\frac{2\sin^2\left(\frac{\nu t}{2}\right)\coth\left(\frac{\beta\nu}{2}\right)+i\sin\left(\nu t\right)}{(\Omega-\nu)^2+\frac{\gamma^2}{4}}\right].
\end{multline}
From this expression we can then evaluate the Fourier transform in
  Eq.~(\ref{eqn:Gdelta}), and thus find the rates
  $\Gamma(\pm\delta_m)$ determining emission and absorption into
  various photon modes, as well as the corresponding Lamb shifts.  The
  appearance of a thermal photon distribution can be traced back to
  properties of these rates, and thus of the correlator given in
  Eq.~\eqref{eqn:DtD0}.  Specifically, thermalization requires that 
  the correlation function obeys the Kubo-Martin-Schwinger
  relation~\cite{Kubo1957, *Martin1959},
  $\average{\hat{D}^\dagger_i(t)\hat{D}^{}_i(0)}=\average{\hat{D}_i^\dagger(-t-i\beta)\hat{D}^{}_i(0)}$.
  If $\Gamma_\uparrow, \Gamma_\downarrow$ could be neglected,
  substituting this into Eq.~(\ref{eqn:Gdelta}) would lead directly to
  the Kennard-Stepanov relation~\cite{Kennard1918, *Kennard1926,
    *Stepanov1957} between emission and absorption rates,
  $\Gamma(\delta)={\rm e}^{\beta\delta}\Gamma(-\delta)$.  However, the expression in Eq.~(\ref{eqn:Gdelta}) involves the convolution of this spectrum with a Lorentzian due to the pump and decay, and so the Kennard-Stepanov relation only holds at small detunings.
  As the detuning is increased
  $\Gamma(\delta)$ ceases to obey this relation because the tails of
  $\Gamma(\delta)$ arise from the Lorentzian broadening with width
  $\Gamma_\uparrow+\Gamma_\downarrow$. The noise temperature of this pump term is not in thermal equilibrium with the dye. We therefore model this pump by a white noise (i.e.\ infinite temperature) bath~\cite{Lax2000}. For experimentally realistic parameter values these effects do not cause any significant deviation from the Kennard-Stepanov relation.

  At this point we note that had we instead used the quantum
  regression theorem to evaluate the correlator
  $\average{\hat{D}^\dagger_i(t)\hat{D}^{}_i(0)}$, the resulting expression would
  not have obeyed the Kennard-Stepanov relation.  The quantum
  regression theorem is known to be incapable of describing finite
  temperature fluctuation dissipation
  relations~\cite{Ford1996,Lax2000}.  The results of the quantum
  regression calculation would correspond to replacing $\nu\to
  \Omega$ in the Keldysh component of the Green's function in Eq.~\eqref{eqn:inverseGF} i.e.\
  assuming that the bath occupation can be represented by sampling it
  at a single frequency $\nu=\Omega$.

  As discussed previously in our Letter, it is the existence of the
  Kennard-Stepanov relation that causes the thermal state of the
  rovibrational degrees of freedom to be imprinted on the photon
  distribution. The thermal spectrum comes about not because the
  emission is thermal, but because the ratio of emission to absorption
  is thermally weighted.  As noted above, this breaks down in the
  tails of the molecular spectrum.  Note that in what follows,
  we consider cases where the lowest cavity mode is detuned below 
  the peak of the molecular spectrum, so that it is typically the lowest
  energy photon modes that fall in the tail of the spectrum, while
  the ``thermal tail'' of the photon distribution is near the
  center of the molecular spectrum.  In addition to relying
  on the thermal nature of the spectrum, such a mechanism of
  thermalization relies on the possibility of light being re-absorbed
  by the dye molecules before it escapes from the cavity.
  This also breaks down in the tails of the spectrum, where
  the absorption and emission rates become small.
  As we will
  discuss below, these points can be clearly seen in the time
  evolution towards a thermal distribution, and in the ways that the
  thermal distribution breaks down.

The quantum model we consider is thus fully defined by
  Eq.~(\ref{eqn:MEK}) along with the definitions of the rates via
  Eqs.~\eqref{eqn:Gdelta} and \eqref{eqn:DtD0}.  We summarize the
  parameters appearing in the model, and the values used in this
  manuscript, in table~\ref{tab:parameters}.

\newcommand{\spread}[1]{\multicolumn{2}{l|}{#1}}

\begin{table*}
  \centering
  
  \begin{tabular}{|c|ll|}
    \hline
    \bf Parameter & \bf Meaning & \bf Value(s) used \\
    \hline
    $\delta_0$ & Lowest cavity mode detuning & $-300$THz to $-100$THz \\
    $\epsilon$ & Cavity mode spacing & $5$THz \\
    $\kappa$ & Cavity mode decay rate &$100$MHz -- $1$THz \\
    $g$ & Light-matter coupling strength&  $1$GHz\\
    $\Gamma_\downarrow$ & Decay rate of excited electronic 
    state\footnotemark[1]\footnotetext[1]{Excluding absorption from and emission into cavity modes} & $1$GHz \\
    $\Gamma_\uparrow$ & Pumping rate of electronic states\footnotemark[1] &
    Variable\\
    $N$ & Number of molecules & $10^9$ \\
    $\Omega_j$ &  Frequency of $j$th rovibrational mode & $5$THz - $60$THz\\
    $\gamma_j$ & Relaxation (thermalization) rate of mode $j$ & $5$THz - $50$THz \\
    $S_j$ & Huang-Rhys factor of mode $j$ & $0.5$ \\
    \hline
    $\Gamma(-\delta_m)$ & \spread{Emission rate into cavity mode $m$} \\
    $\Gamma(\delta_m)$ & \spread{Absorption rate from cavity mode $m$} \\
    $\tGu,\tGd$ &
    \spread{ Electronic transition rates \emph{including} contribution of cavity modes} \\
    \hline
  \end{tabular}
  \caption{Summary of bare and derived parameters for the quantum model we consider.  For each parameter, the values or range of values used in this manuscript are quoted, as appropriate.}
  \label{tab:parameters}
\end{table*}

\subsection{Multiple vibrational modes}
\label{sec:multimode}

The absorption and emission spectra of dye molecules observed in
experiments~\cite{Schafer1990, Klaers2010c} show a structure with
multiple peaks. There are two possible origins of this feature; either
the relaxation time of the rovibrational states is long enough to
allow multi-phonon effects to be spectrally resolved, or there
are multiple vibrational modes with different frequencies which are
important. The very rapid thermalization of the system seen in
experiments~\cite{Klaers2010c} rules out the first of these options
and so here we consider the second case. To do this we modify the
Hamiltonian in Eq.~\eqref{eqn:H} to include multiple vibrational
modes for each molecule, so that it now reads
\begin{multline}
  \label{eqn:Hmulti}
  \hat{H}=\sum_{\modelabel}\omega_\modelabel\ad_\modelabel \hat{a}_\modelabel
    +  g\sum_{\modelabel,i}\left(\hat{a}_\modelabel\hat{\sigma}_i^++\ad_\modelabel\hat{\sigma}_i^-\right)
  \\
   +\sum_{i}\frac{\omDye}{2}\hat{\sigma}_i^z
  +\sum_j\Omega_j\left[\bd_{i,j}\hat{b}_{i,j}+\sqrt{S_j}\hat{\sigma}_i^z(\hat{b}_{i,j}+\bd_{i,j})\right].
\end{multline}
There is now a sum over $j$ which indexes the rovibrational modes at
each site and so $\bd_{i,j}$ is an operator which creates a
vibrational excitation in mode $j$ of molecule $i$. We can then
go through exactly the same calculation as in the previous
section. This results in a master equation with exactly the same form
as Eq.~\eqref{eqn:ME} but where the displacement operator is now a
product of single mode operators and hence the correlation function
now includes contributions from all of the vibrational modes
\begin{align} 
  \average{\hat{D}^\dagger_i(t)\hat{D}^{}_i(0)}= \exp\left( - \sum_j \frac{2 S_j\gamma_j}{\pi} f_j(t)
  \right),
  \nonumber\\
  f_j(t)= \int_{-\infty}^\infty
  d\nu\frac{2\sin^2\left(\frac{\nu
        t}{2}\right)\coth\left(\frac{\beta\nu}{2}\right)+i\sin\left(\nu
      t\right)}{(\Omega_j-\nu)^2+\frac{\gamma_j^2}{4}}.
\end{align}
We note that the spectrum which results from this expression still obeys the Kennard-Stepanov relation with the same caveats as above.

\begin{figure}
  \centering
  \includegraphics[width=1.6 in]{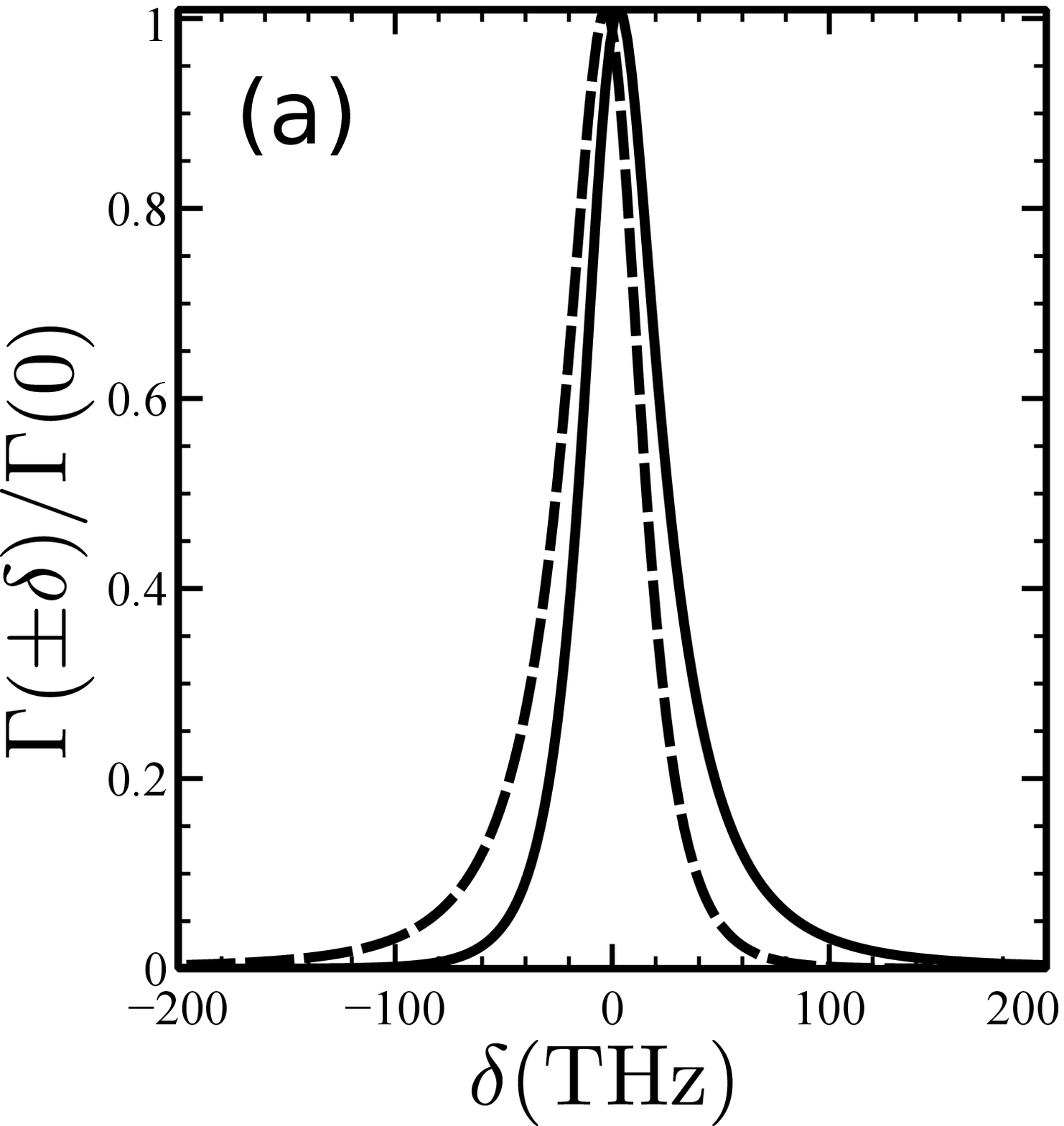}
  \includegraphics[width=1.6 in]{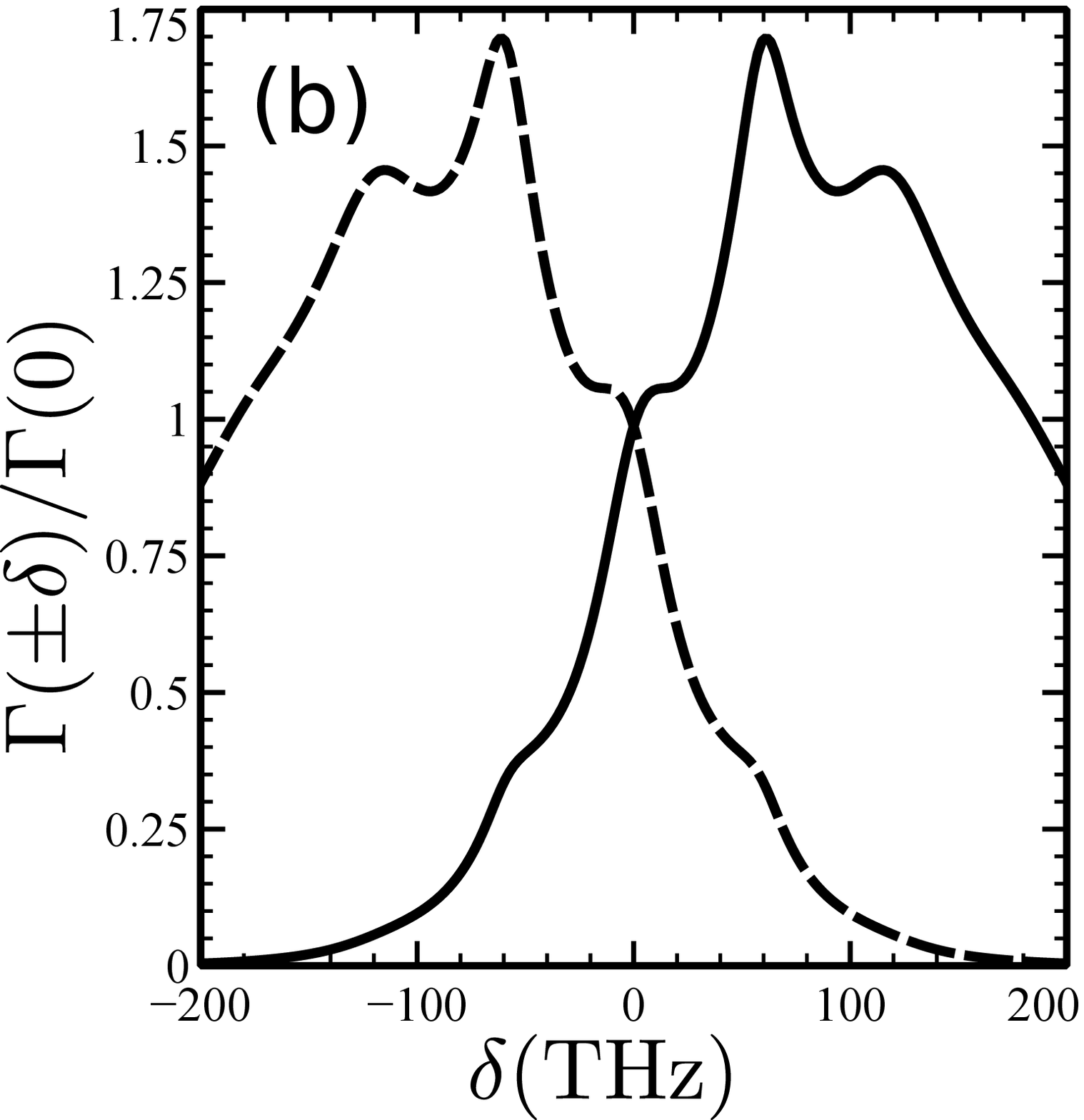}
  \caption{The effective absorption (solid) and emission (dashed) rates, $\Gamma(\pm\delta)$ for different vibrational mode structures. In (a) we show the case of one mode with parameters $S=0.5$, $\Omega=5$THz, $\gamma=50$THz. In (b) we also include a second mode with $S=0.5$, $\Omega=60$THz, $\gamma=30$THz}\label{fig:phonons}
\end{figure}

We show examples of the resulting absorption and emission spectra
including one and two vibrational modes in Fig.~\ref{fig:phonons}. In
the case with only one low frequency mode included the spectrum simply
consists of a single peak broadened by the thermalization rate. The
difference between the location of the maxima in the absorption and
emission scales with the coupling between the vibrational and
electronic degrees of freedom, $S$. When we add an extra higher
frequency mode we see that the spectrum gains multiple peaks at
integer multiples of the second mode frequency and reduces the
relative weight of the spectrum around $\delta=0$ (the zero phonon
line). This type of spectrum is much closer to that seen in the
absorption and emission of the dyes used in the photon condensation
experiments~\cite{Klaers2010c}.

Using this multi-peaked type of spectrum, however, does not affect the
physics which we will go on to discuss. The mechanisms of
thermalization and the way they break down do not depend on the
details of the absorption and emission rates.  However, the
  behavior once thermalization has broken down will depend on the
  shape of the emission spectrum.  For the remainder of this paper we
use the simpler single mode case except where otherwise noted.

\section{Rate equation approximation} \label{sec:rate}

Having derived a full quantum model in the previous section, we
  next turn to study the properties of this model using a rate equation
  approximation. Such an approximation is valid when there are many
  molecules coupled to each photon mode, so that quantum correlations
  are suppressed by $1/N$.  However, such an approximation is
  restricted to describing physics that depends only on the
  populations of the modes, and not on the off-diagonal coherences,
or higher order statistics (such as $g^{(2)}$).  As discussed below, a
closed set of equations for the mode populations exists and this is considerably easier to solve than
Eq.~(\ref{eqn:ME}).  We will however return in
Section~\ref{sec:linewidths} to discuss features beyond the scope of
the rate equation.

\subsection{Derivation of rate equation}
\label{sec:deriv-rate-equat}

To obtain the rate equation we write down an equation for the
expectation value of the number of photons in a given mode
$n_m=\average{\ad_m \hat a^{}_m}$ and make the semiclassical
approximation that the density operator for the photons and molecules
factorizes e.g.\ $\average{\ad_m \hat a^{}_m \hat
  \sigma^+}=\average{\ad_m \hat a^{}_m}\average{\hat \sigma^+}$. This
then leads to the set of coupled equations for the evolution of
the populations of the photon modes, $n_m$, and for the
  probability of finding a molecule in its excited state, $p_e$. For
$N$ dye molecules these are given by,
\begin{gather} \label{eqn:nmdot}
\begin{split}
	\pardiff{n_m}{t}=-\kappa n_m+N\left[\Gamma(-\delta_m)(n_m+1)p_e  \right.\\
	\left.-\Gamma(\delta_m) n_m(1-p_e)\right],
\end{split}  \\
	\label{eqn:pedot}\pardiff{p_e}{t}=-\tGdn p_e+ \tGun(1-p_e),
\end{gather}
where we have defined the rates:
\begin{gather}
	\tGun=\Gu+\sum_{\modelabel}g_{\modelabel}\Gamma(\delta_{\modelabel})\nl, \\
        \tGdn=\Gd+\sum_{\modelabel}g_{\modelabel}\Gamma(-\delta_{\modelabel})(\nl+1).
\end{gather}
  We see from the expressions above, that the rate of emission and
  absorption into a given cavity mode depend on the number of excited
  state molecules, and on the number of photons already in that mode,
  exactly as one would expect.  Similarly, the transition rates
  between the electronic states depend on the numbers of photons in
  all modes.  As we will see below, this has the consequence that the
  populations of the modes are coupled, and leads to the emergence of a
  chemical potential for photons.

\subsubsection{Steady state distribution}
\label{sec:steady-state-distr}

If we are only interested in the steady state properties of the photon distribution, then we may adiabatically eliminate the molecular degrees of freedom and obtain the self-consistent expression for the photon distribution
\begin{equation} \label{eqn:dndt}
  \kappa\nl=N\frac{\Gamma(-\delta_\modelabel)(\nl+1)\tGu-\Gamma(\delta_\modelabel)\nl\tGd}{\tGu+\tGd}.
\end{equation}
In the equilibrium limit when the losses from the cavity are negligible,  $\kappa, \Gd, \Gu \to 0$, this expression results in a Bose-Einstein distribution for the photons which satisfies
\begin{equation}
	\frac{n_m+1}{n_m} = e^{\beta \delta_m} \frac{\tGd}{\tGu}.
\end{equation}
 We are thus
able to define an effective chemical potential $\mu=k_B T \ln
{\tGu/\tGd}$ which, far below threshold, when the populations of all
the photon modes are negligible, can be approximated as $\mu_0=k_B T
\ln {\tGu(0)/\tGd(0)}$.  NB, while the effective pumping rate for
  an empty cavity $\tGu(0)= \Gu$, the effective decay rate $\tGd(0)
  \neq \Gd$, due to existence of spontaneous emission into the cavity
  modes.

\begin{figure}
  \centering
  \includegraphics[width=3.4 in]{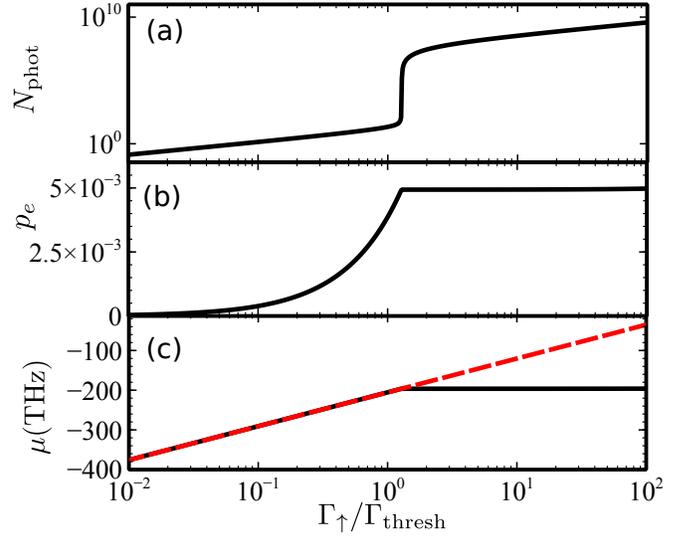}
  \caption{(Color online) Steady state behavior of our
     model as it is pumped through the BEC transition. (a)
    shows the total number of photons in the cavity. (b) gives the
    fraction of molecules in the excited state and (c) shows the
    effective chemical potential. In (c) we also plot (as the red,
    dashed curve) the simple below threshold expression
    $\mu_0=k_BT\ln {\tGu(0)/\tGd(0)}$. The parameters used are $S=0.5$,
    $\gamma=50$THz, $\Omega=5$THz, $\kappa=100$MHz, $T=300$K,
    $N=10^9$, $g=1$GHz, $\delta_0=-200$THz,
    $\Gd=1$GHz.}\label{fig:Ntotinvmu}
\end{figure}

In Fig.~\ref{fig:Ntotinvmu} we show the behavior of the system as we
increase the pump strength through this threshold. We see that the
total number of photons in the cavity $N_{\rm phot}=\sum_mg_m n_m$ has
a sharp transition at the same point as the number of excited
molecules saturates. This also corresponds to the chemical potential
reaching the energy of the ground mode of the cavity. We note that the
behavior of both $N_{\rm phot}$ and $p_e$ is qualitatively what is expected
for the transition to a lasing state, but the saturated value for $p_e$ is much lower than the electronic inversion point. Also note that in a laser it is impossible
to identify a chemical potential.    In Fig.~\ref{fig:Ntotinvmu}(c) the
chemical potential defined in this way shows exactly the same behavior
as is typically associated with a BEC transition.  Below threshold we
see that the chemical potential matches very well to the simple
expression derived above assuming the cavity is empty; at and
  above threshold, the chemical potential locks to $\mu=\delta_0$.

\subsubsection{Threshold condition}
\label{sec:threshold-condition}

Before discussing the phase boundary of the photon condensate
  state, we must first define the threshold condition carefully.  Care
  is required because we consider a finite system in an harmonic trap,
  and so the boundary is not sharp.  We wish to define a threshold
  condition in terms of the number of photons in the ground mode of
  the cavity, $n_0$.  A naive approach would be to define the
  threshold when this reaches a fixed value, e.g.\ $1$, however if
  applied to the equilibrium limit, this gives a complicated formula
  for the critical temperature.  In equilibrium, the appropriately
  defined thermodynamic limit~\cite{Bagnato1991} has a critical
  temperature $k_B T = \epsilon\sqrt{6 N_{\text{phot}}}  /\pi$, where
  $N_\text{phot}$ is the total number of photons in the cavity.  We
  aim to define the threshold condition consistently with this.

We begin by assuming the system is in thermal equilibrium. For our 2D harmonically trapped gas, the energy of the mode with $n$ and $m$ photons in the two directions and mode spacing $\epsilon$ is given by $\delta_0+(n+m)\epsilon$. The total number of particles
in the Bose-Einstein distribution is therefore
\begin{gather}
	N_{\rm phot}= \sum_{n,m} \left[{\rm e}^{\beta [(n+m)\epsilon - \mu+\delta_0]} -1\right]^{-1}\nonumber \\
	=\sum_{j=1}^\infty \left(1-{\rm e}^{-\beta\epsilon j}\right)^{-2}{\rm e}^{\beta (\mu-\delta_0) j}.
\end{gather}
In the thermodynamic limit, $\epsilon \to 0$, the transition occurs
exactly at $\mu=\delta_0$, and so ${\rm e}^{\beta (\mu-\delta_0)}=1$.  To leading order
in $\beta \epsilon$ this gives the critical photon number:
\begin{equation}\label{eqn:totpart}
	N_{\rm phot}=\frac{(k_BT)^2}{\epsilon^2} \sum_{j=1}^\infty \frac{1}{j^2}
        =
        \frac{(k_B T)^2 \pi^2}{6 \epsilon^2},
\end{equation}
which diverges as $\epsilon \to 0$.  Even at non-vanishing
  $\epsilon$, if we were to define the threshold condition as
$\mu=\delta_0$, we would find a ground mode population
$n_0\rightarrow\infty$. As such, we cannot use $\mu=\delta_0$ as the
threshold condition. To regularize this we calculate the
next-to-leading order contribution to the total particle number in
$\beta\epsilon$. This is given by
\begin{gather}
  N_{\rm phot} \simeq\frac{1}{(\beta\epsilon)^2}\sum_{j=1}^\infty
  \frac{{\rm e}^{\beta (\mu-\delta_0) j}}{j^2\left(1-\beta\epsilon\right)^j},
\end{gather}
where we have used $1-j\beta \epsilon=(1-\beta \epsilon)^j$ to order
$\beta\epsilon$.  If we define the threshold as occurring when
  $N_{\text{phot}}$ reaches the value given in
  Eq.~(\ref{eqn:totpart}), then we must
choose~\cite{Bagnato1991,Hadzibabic2008, Kirton2013b}
\begin{equation}
  e^{\beta (\mu-\delta_0)} = 1 -\beta \epsilon 
  \qquad \to \qquad
  {n_0}=\frac{1}{\beta\epsilon} - 1.
\end{equation}
To leading order in $\epsilon$, we thus use $n_0=1/\beta \epsilon$
  to define the threshold condition. From this we can go on to define
a threshold pump power, $\Gamma_{\rm thresh}$ as the value of $\Gu$
required for this population to be reached.  Note that since
$N_\text{phot} \sim 1/(\beta \epsilon)^2$, the criterion we use
corresponds to $n_0 \sim \sqrt{N_{phot}}$, and so can be understood as
distinguishing macroscopic ($n_0 \sim N_{\text{phot}}$) and
microscopic ($n_0 \sim 1$) occupations of a single mode.

Away from thermal equilibrium, when the system does not obey a
Bose-Einstein distribution, it will still be useful to define the same
threshold but in this case we will use the slightly more general
definition $\max\{n_m\}=1/\beta\epsilon$. That is, whenever one mode of
the cavity exceeds the required population.

\subsection{Breakdown of thermalization}  \label{sec:breakdown}

We have shown that, for weak enough losses, the master equation given
in Eq.~\eqref{eqn:ME}, has as its steady state solution an equilibrium
Bose-Einstein distribution.  As such, the steady state of this
equation shows a condensation transition at a critical photon density,
or equivalently a critical pump power. To reach this thermal
equilibrium state we have shown that it is necessary for the dye
molecules to thermalize with the photon distribution. Hence, the
timescale over which the thermalization happens must be shorter than
the lifetime of photons inside the cavity. In our earlier
Letter~\cite{Kirton2013b} we showed how the distribution crosses over
to that of a standard laser as the cavity lifetime is decreased.  In
this manuscript we instead concentrate on varying other properties of
the cavity, and the vibrational properties of the dye molecules.

\subsubsection{Changing the cavity cutoff}

We begin by considering what happens when the length of the cavity is
changed, as was experimentally~\cite{Klaers2010c} tested. When the
length of the cavity is increased the energy of the lowest frequency
mode it can support decreases. The detuning of these low energy
modes from the dye molecule resonance then increases, and so the
modes have very small absorption and emission rates
$\Gamma(\pm\delta_0)$.  Thus, for these modes, cavity losses compete
with absorption and emission.

\begin{figure}
  \centering
  \includegraphics[width=3.3 in]{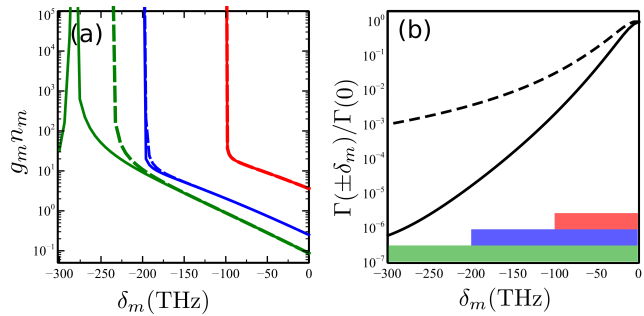}
  \caption{(Color online) (a) Photon distribution just above threshold pump
    power, illustrating the effect of decreasing the energy of the
    lowest cavity mode, $\delta_0$. The green curve is
    $\delta_0=-300$THz, blue is $\delta_0=-200$THz, and red is
    $\delta_0=-100$THz. The dashed lines show Bose-Einstein fits to
    the tails of the data. (b) The absorption (solid) and emission
    (dashed) rates for the same parameters. The shaded regions show
    which modes are included for the same colored curves as in
    (a). All parameters except $\delta_0$ are the same as in
    Fig.~\ref{fig:Ntotinvmu}.}\label{fig:varyd0}
\end{figure}

We show the effects of decreasing $\delta_0$ on the system in
Fig.~\ref{fig:varyd0}(a) by comparing the numerical steady state
distributions (solid lines) to Bose-Einstein distribution fits to the
tail of the numerics (dashed lines).  In order to perform this fit we
first fix the temperature to be that of the dye, and then use the
chemical potential as a variable parameter to fit to the thermal tail
of the numerical results.  As can be seen, for the parameters used in
Fig.~\ref{fig:varyd0}(a) a thermal tail with the correct temperature
is always present (since this thermal tail is near the center of the
molecular spectrum, $\delta=0$, thermalization is good).  Adjusting
the chemical potential corresponds to fitting an overall scale for the
intensity of the thermal tail, thus this can be fit by matching a
single point in the tail of the distribution.  Since such a fit is
matched only to the higher energy photon modes, there is no guarantee
as to how the extracted chemical potential $\mu$ compares to the
lowest energy photon mode energy $\delta_0$.  In equilibrium,
$\mu\le\delta_0$ with equality holding above threshold, when the
chemical potential locks to the bottom of the spectrum.  Since
  $\mu$ is extracted from the high energy tail, whether or not it
  matches the low energy peak and thus locks at the cutoff frequency
  $\delta_0$ can provide a good measure of the degree of
thermalization of the distribution, as discussed further below.

When the cutoff is $-100$THz (the red curve) we see that the system
reaches thermal equilibrium and the photon populations are well
described by a Bose-Einstein distribution. Reducing $\delta_0$ to
$-200$THz (the blue curve) we see that the match to a Bose-Einstein
distribution is still good but there is a slight discrepancy in the
prediction of the location of the peak calculated just from looking at
the thermal tail.  This disagreement between the numerical results and
an equilibrium distribution is even more apparent in the curve with a
detuning of $-300$THz (the green curve) where the lowest energy modes
are completely out of equilibrium and the macroscopically occupied
mode is one of the excited modes of the cavity.  As discussed above,
this breakdown is due to the cavity losses being too fast for these
modes with low absorption and emission rates to thermalize.  This is
the same mechanism as discussed in our previous
work~\cite{Kirton2013b}, where we considered the effect of reducing
the cavity lifetime. We note that the reason that the thermal tails
  of the photon distributions are not exactly parallel is
that these curves are Bose-Einstein distributions multiplied by a
degeneracy factor which results in logarithmic corrections to the tail.
These become more important for smaller cutoff energies.

\begin{figure}
  \centering
  \includegraphics[width=1.65 in]{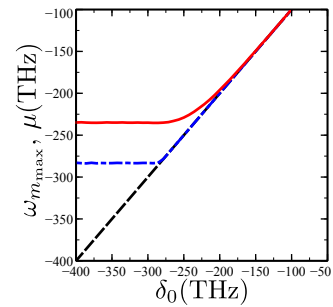}
  \caption{(Color online) Comparison between fitted
      chemical potential, energy of maximally populated mode and
      cavity cutoff, as the cutoff is varied.  Above threshold in
      equilibrium, the maximally occupied mode and effective chemical
      potential both lock to the cavity cutoff (dashed line).  The
      solid (red) and dot-dash (blue) lines correspond, respectively, to
      the fitted chemical potential, and energy of the maximally
      populated mode as extracted from the steady state distribution
      for the same parameters as given in
      Fig.~\ref{fig:Ntotinvmu}.}\label{fig:varyd0mu}
\end{figure}

In equilibrium, above threshold, both the maximum in the photon
distribution, $\ommax$, and the value of the chemical potential, $\mu$
(found from the Bose-Einstein fit to the tail) lock at the energy of the ground mode
of the cavity, $\delta_0$.  As such, the difference between these
quantities and the ground mode energy can be used to demonstrate the
breakdown of thermalization.  In Fig.~\ref{fig:varyd0mu} we show the
way in which these three quantities vary as we change the cavity
cutoff energy.  At small values of $\delta_0$ these all match the
equilibrium expectation (the energy of the cavity ground mode).  As
the detuning is increased the first change is that the fitted chemical
potential starts to deviate, while the macroscopically occupied mode
of the cavity remains the cavity ground mode.  This situation
corresponds to that seen in the blue curve of
Fig.~\ref{fig:varyd0}(a), where a slight deviation from a thermal
distribution is visible.  As the detuning is increased even further,
the ground mode of the cavity becomes far detuned from the molecular
emission peak, and so the rates of absorption and emission can no
longer compete with cavity losses.  One then has that an excited mode
of the cavity gains a macroscopic occupation and so the energy of this
maximally populated mode deviates from the equilibrium prediction.  It
is notable that at the smallest values of $\delta_0$, both the fitted
chemical potential and the energy of the maximally populated
modes saturate.  This occurs because the additional low energy cavity
modes have negligible population (as cavity loss beats emission rate),
and so the behavior of the system is not affected by including these
extra low energy modes.
\begin{figure}
  \centering
  \includegraphics[width=2.5 in]{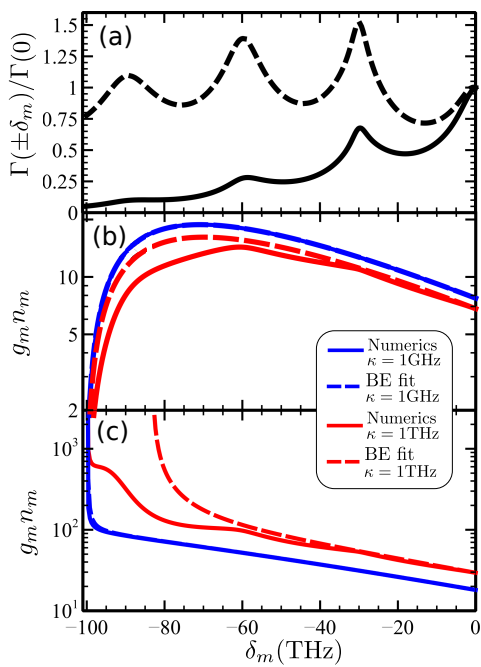}
  \caption{(Color online) The effect of a peaked absorption and
    emission spectrum. In (a) we show the form of
    $\Gamma(\pm\delta_m)$ when we included a vibrational mode which is
    not overdamped (solid line shows absorption, dashed line
    emission).  In (b) and (c) we show the effect of increasing the
    cavity losses for this spectrum below ($\Gu=0.5\Gamma_{\rm thresh}$) and above ($\Gu=2.5\Gamma_{\rm thresh}$) threshold
    respectively. In each case we plot the behavior for two cavity
    loss rates: $\kappa=1$THz (red) and $\kappa=1$GHz (blue).  The
    dashed lines are the fit to a Bose-Einstein distribution. Other
    parameter values are the same as Fig.~\ref{fig:Ntotinvmu} except
    as follows: The two vibrational modes are characterized by
    $S_1=0.1$, $\Omega_1=5$THz, $\gamma_1=50$THz, $S_2=0.5$,
    $\Omega_2=30$THz, $\gamma_2=5$THz.  }\label{fig:thermalholes}
\end{figure}

\subsubsection{Changing the properties of the dye molecules} 

By changing the properties of the dye molecules it is possible to
change the functional form of $\Gamma(\delta)$.  In particular, by
  introducing coupling to extra vibrational modes, as discussed in
  Sec.~\ref{sec:multimode}, it is possible to engineer a form for
  $\Gamma(\delta)$ which has multiple peaks.  To achieve such a
  multi-peaked structure, one needs spectrally resolved vibrational
  sidebands, which requires that (some of) the vibrational modes must
  have frequencies larger than their linewidths, i.e.\ be underdamped.
An example of this type of spectrum is shown in
Fig.~\ref{fig:thermalholes}(a). This is similar to the type of
spectrum seen experimentally~\cite{Klaers2010c} but with a more
exaggerated multi-peaked structure. We note that while the
spectrum now looks very different to the one used in the rest of this
paper it still obeys the Kennard-Stepanov relation and so in thermal
equilibrium will give rise to a Bose-Einstein distribution for the
photons. We see that this is the case in Fig.~\ref{fig:thermalholes}(b) and (c) where for small cavity losses (the blue solid curve) the
numerical results match well with the Bose-Einstein fit both above and
below threshold. As the losses from the cavity are increased we see
that the distribution becomes non-thermal, but it does so in a more
complicated way than when the absorption and emission spectra have only a
single peak. In this case, the modes close to the minima in
$\Gamma(\delta)$ are the ones which are no longer able to thermalize,
since the absorption and emission rates here are small enough that the
cavity lifetime is too short for thermal equilibrium to be
reached. This causes the complex non-monotonic photon spectra seen in
the solid red curves of Fig.~\ref{fig:thermalholes} (b) and (c) which
now significantly deviate from the equilibrium fit shown by the
dashed curve.

The results above, which show how it is possible to break the
thermalization process, give motivation to the choice of two possible
criteria for thermalization which characterize the behavior of the
system at or near threshold and determine whether it is in thermal
equilibrium. Firstly, we can look at which mode of the cavity gains a
macroscopic occupation, $\ommax$, when we pump the system above
threshold. When this mode is the ground mode of the cavity the system
is close to thermal equilibrium but when this mode is one of the
excited cavity the distribution has failed to thermalize. Secondly, we
can look at the chemical potential of a fit to the distribution at or
above threshold. If this is close to the ground mode energy then the
system is in thermal equilibrium. These allow us to draw ``phase
diagrams'' which separate regions in which the system looks like a
thermal equilibrium condensate from those where it is more like the
non-thermal state of a laser. The algorithm for generating
these plots is as follows: for each set of parameter values the pump
power is increased until the threshold (as described in
Sec.~\ref{sec:threshold-condition}) is reached, then the number of the
mode with the largest population and the chemical potential of the fit
to the tail of the thermal distribution are recorded.

\begin{figure}
  \centering
  \includegraphics[width=1.65 in]{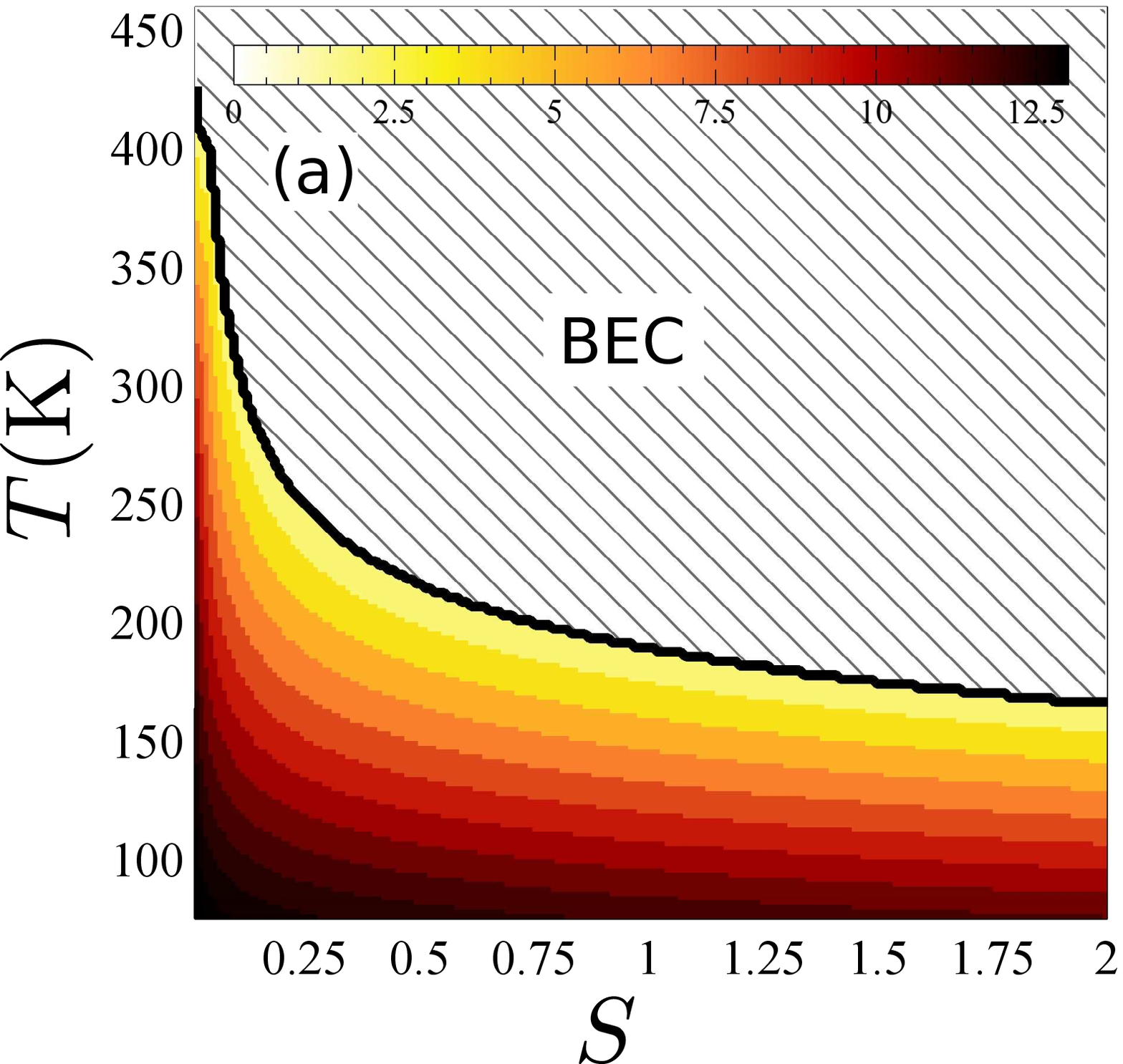}
  \includegraphics[width=1.65 in]{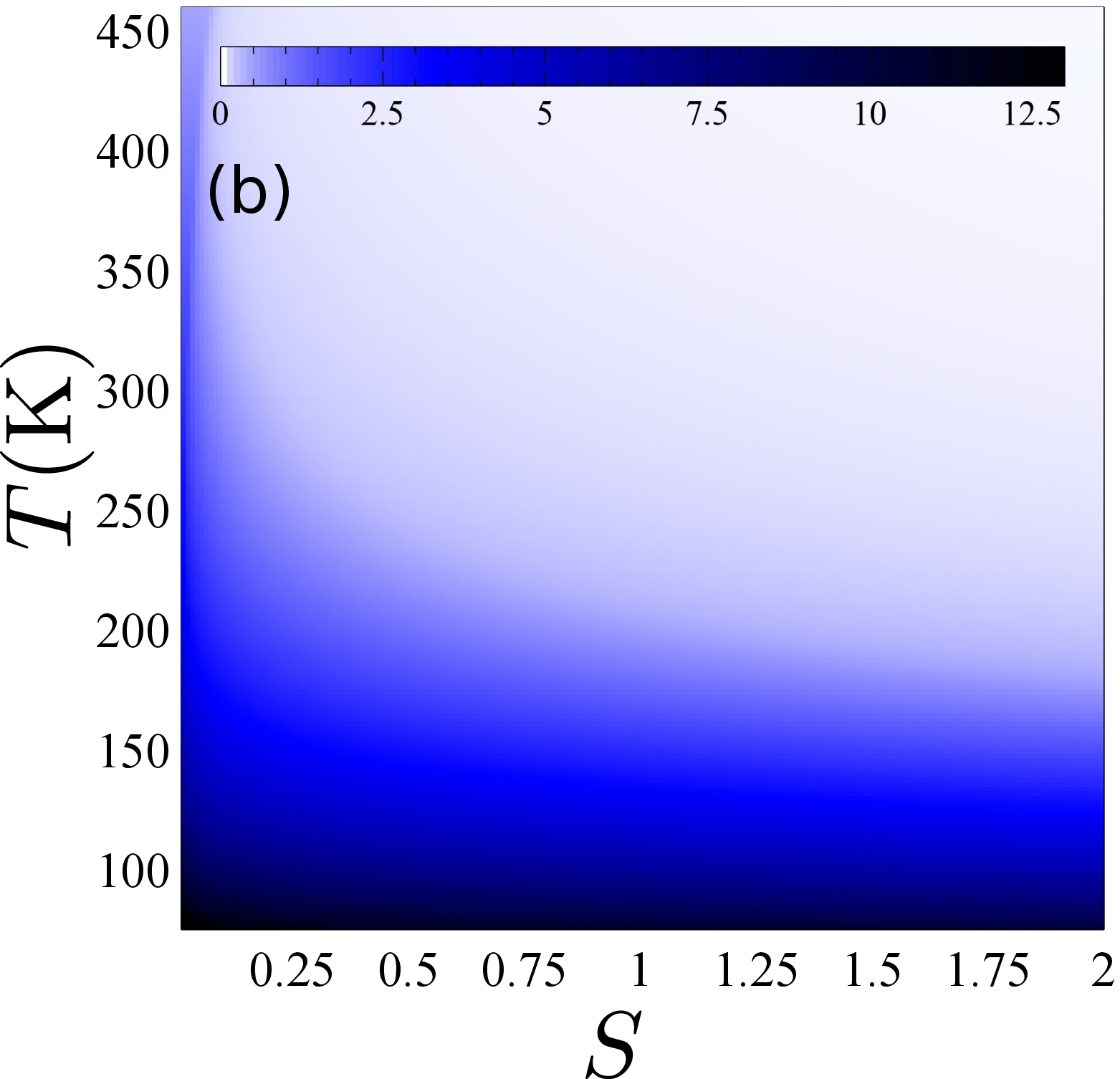}
  \caption{(Color online) Criteria for thermalization vs.\ $S$ and
      $T$. In (a) the color shows the index of the mode which gains a
    macroscopic occupation, the solid line separates the region where
    the ground mode condenses (hashed) from the region where excited
    cavity modes are macroscopically occupied. In (b) we show a
    similar diagram but with the `order parameter'
    $\beta(\mu-\delta_0)$.}\label{fig:SvsT}
\end{figure}

We begin by examining how varying the strength of the coupling
between electronic and vibrational states of the dye molecules,
$S$, can affect the thermalization process. The Huang-Rhys parameter, $S$, describes the difference in displacement between the lowest energy vibrational state in the ground electronic manifold and the lowest energy state in the excited electronic manifold in units of the harmonic oscillator length. In
Fig.~\ref{fig:SvsT}, we look at the behavior of the two criteria
described above in the $S$ vs.\ $T$ plane. For the chemical potential we
plot the dimensionless quantity $\beta(\mu-\delta_0)$ which is zero in
thermal equilibrium and becomes more positive as the thermalization
breaks down.

In the limit $S\to 0$ the absorption and emission spectra are exactly
symmetric and therefore equal, and so thermalization is never possible. Thus, the
lasing--condensation crossover as identified by both the criteria
discussed above moves to infinite temperature. As $S$ is increased,
the asymmetry in the rates increases, and so the minimum temperature
at which thermalization occurs decreases.  We see however that there
is a region where the lowest mode is macroscopically occupied, but the
chemical potential fit to the tail deviates from this lowest mode.
This is the same behavior as was seen in Fig.~\ref{fig:varyd0}(b):
macroscopic occupation of the lowest mode is a weaker criterion.
As the temperature or $S$ is decreased the value of $\beta(\mu-\delta_0)$ increases. As the system is brought further from thermal equilibrium the chemical potential fit to the tail of the distribution moves closer to the gain maximum of the dye.

\begin{figure}
  \centering
  \includegraphics[width=1.65 in]{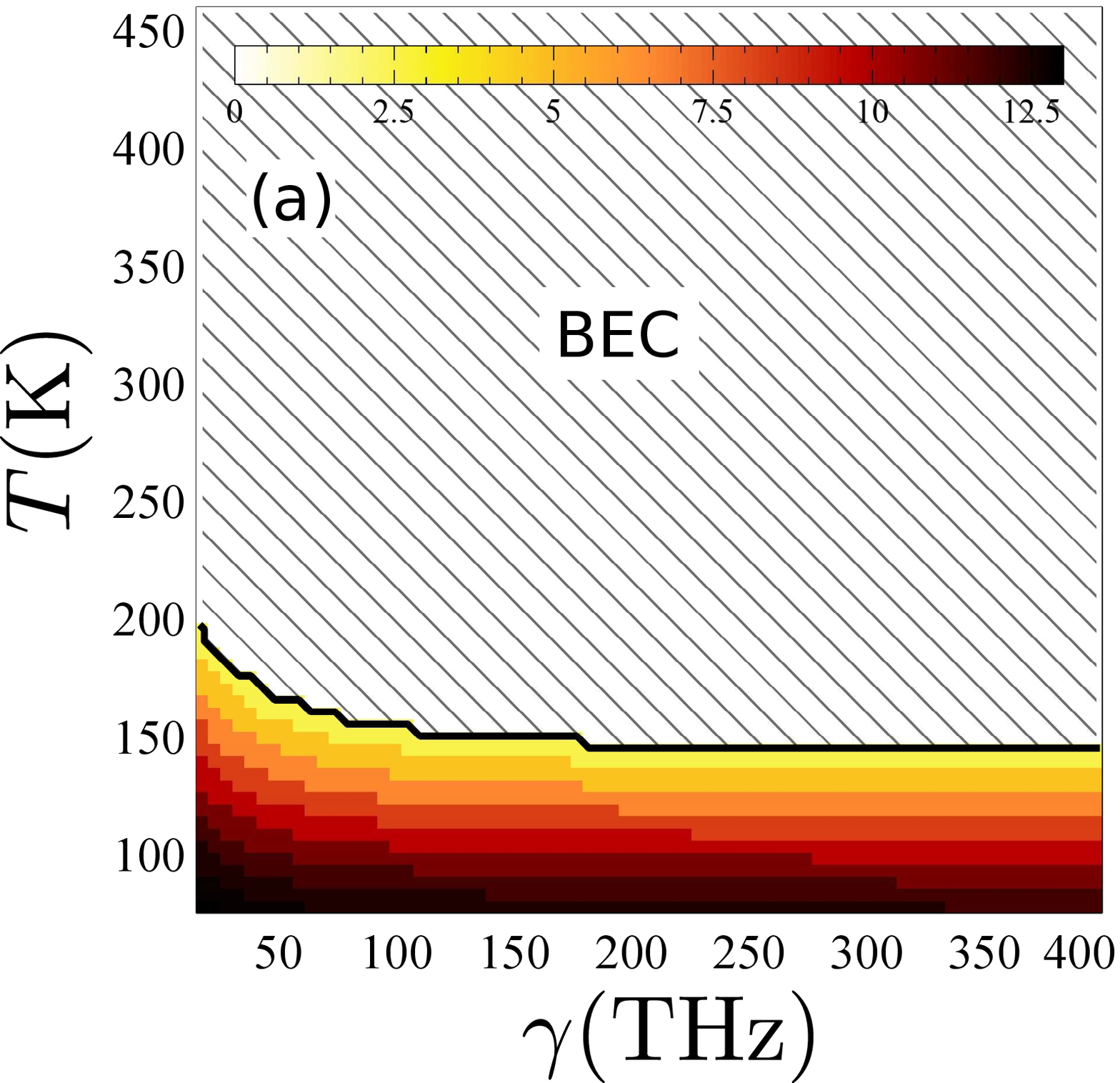}
   \includegraphics[width=1.65 in]{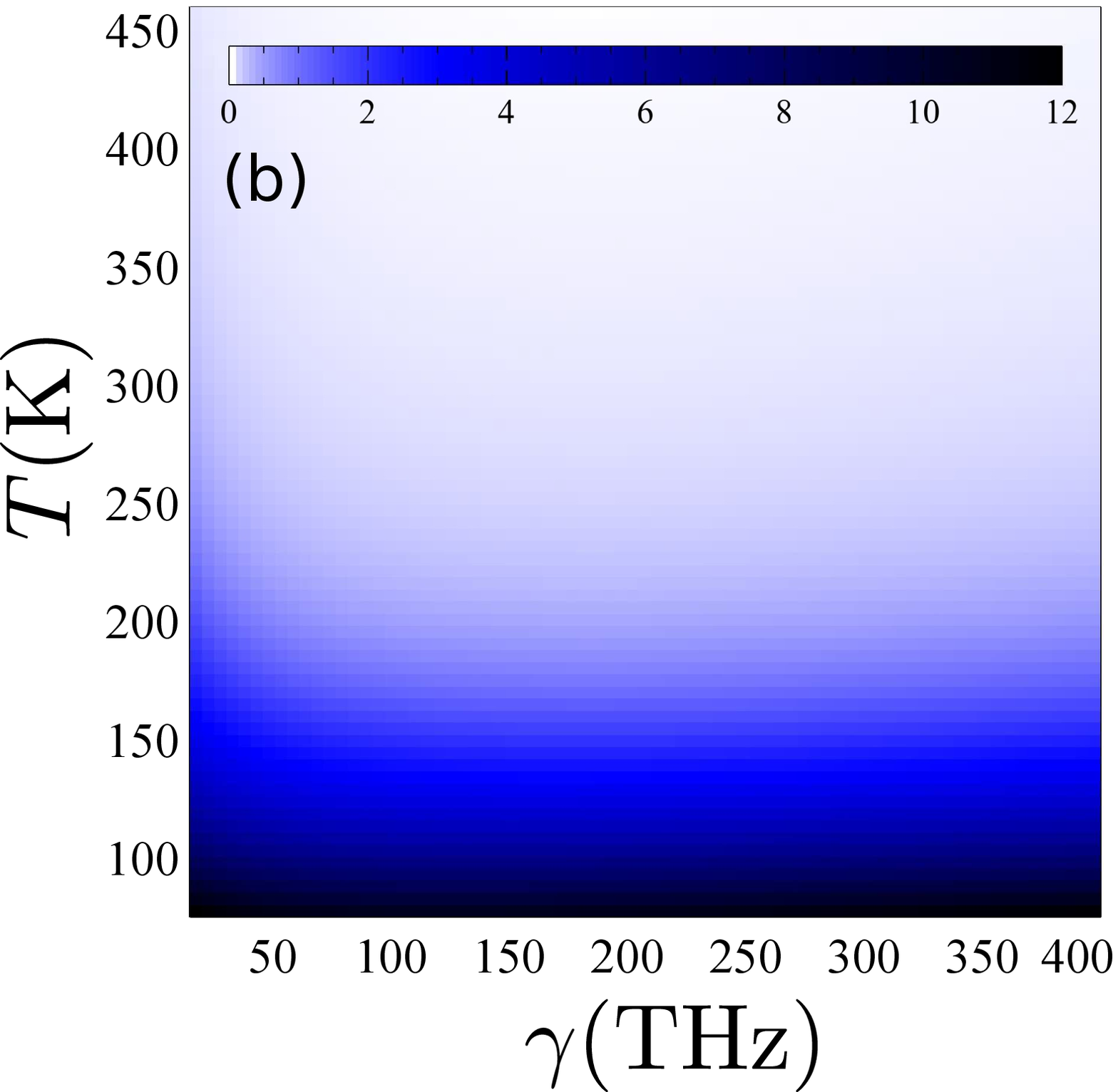}
   \caption{(Color online) Criteria for thermalization vs.\ $\gamma$
       and $T$. Panel (a) shows the index of the mode which is
     macroscopically occupied above threshold. Panel (b) shows the
     chemical potential of the Bose-Einstein fit to the data,
     $\beta(\mu-\delta_0)$. }\label{fig:gvsT}
\end{figure}

We can also look at what happens when we vary the thermalization rate
of the vibrational mode of the molecules, $\gamma$. The phase diagrams
for these results in the $\gamma$ vs.\ $T$ plane are shown in
Fig.~\ref{fig:gvsT}.  At small $\gamma$ the behavior is very similar
to that seen at small values of $S$ above. This is to be
  anticipated, given the form of Eq.~(\ref{eqn:DtD0}) where the
combination $S \gamma$ appears as a prefactor in the exponent.  Thus
the $S\to 0$ and $\gamma \to 0$ limits are similar.

At large $\gamma$ a different behavior occurs due to the Lorentzian
broadening of the vibrational resonances corresponding to the
$\gamma^2$ term in the denominator.  This broadening means the
spectral weight, and thus both the absorption and emission rates, at
any one frequency is suppressed.  This has the consequence that cavity
losses start to compete with absorption and emission, and
thermalization breaks down.  In contrast to the breakdown of
thermalization at small $S$, small $\gamma$, or low temperature, the
breakdown of thermalization at large $\gamma$ occurs simultaneously
across the whole spectrum.  i.e., rather than the just low energy
modes becoming decoupled, all modes cease to follow a thermal
distribution at once.  Furthermore, because the breakdown of
thermalization is not specific to the low energy modes, the
macroscopically occupied mode remains the ground mode. This can be
seen in Fig.~\ref{fig:gvsT} where with increasing $\gamma$ at
  large $\gamma$, the criterion for thermalization given by the
fitted chemical potential moves to higher temperature, while the
criterion given by which mode is occupied continues to move to lower
temperatures.
  
\begin{figure}
  \centering
   \includegraphics[width=3.2 in]{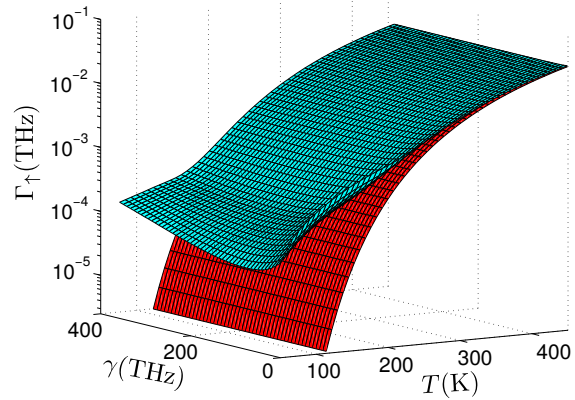}
   \caption{(Color online) Phase diagram for condensation/lasing, showing the
     required pump power for a mode to go above threshold (blue)
     compared to the equilibrium prediction
     (red). }\label{fig:threshvsgT}
\end{figure}
 
A further way of quantifying the deviation from thermalization
can be found be looking at the pumping strength required for the
population in one of the modes to exceed the threshold population
described in Sec.~\ref{sec:threshold-condition}. Comparing the value
of $\Gu$ necessary to exceed this threshold in $\maxnm$ to the
predicted equilibrium value, $\Gamma_{\rm
  thresh}=\exp(\beta\delta_0)\Gd$, we see very similar behavior to the
``phase diagram'' for $\beta(\mu-\delta_0)$. The actual threshold
(blue) is always greater than the equilibrium prediction (red): the
losses always need to be compensated. At low temperatures and small
$\gamma$ we see the calculated threshold rise, this is as the lasing
mode moves to higher excited states of the cavity as explained
previously. We see that at large values of $\gamma$ there is also a
rise in the threshold even in regions where the ground mode is maximally
occupied; this is a signature of the thermalization process breaking
down across the whole spectrum of the cavity.

\subsection{Dynamics of thermalization}
\label{sec:dynam-therm}

\begin{figure}
  \centering
  \includegraphics[width=3.1 in]{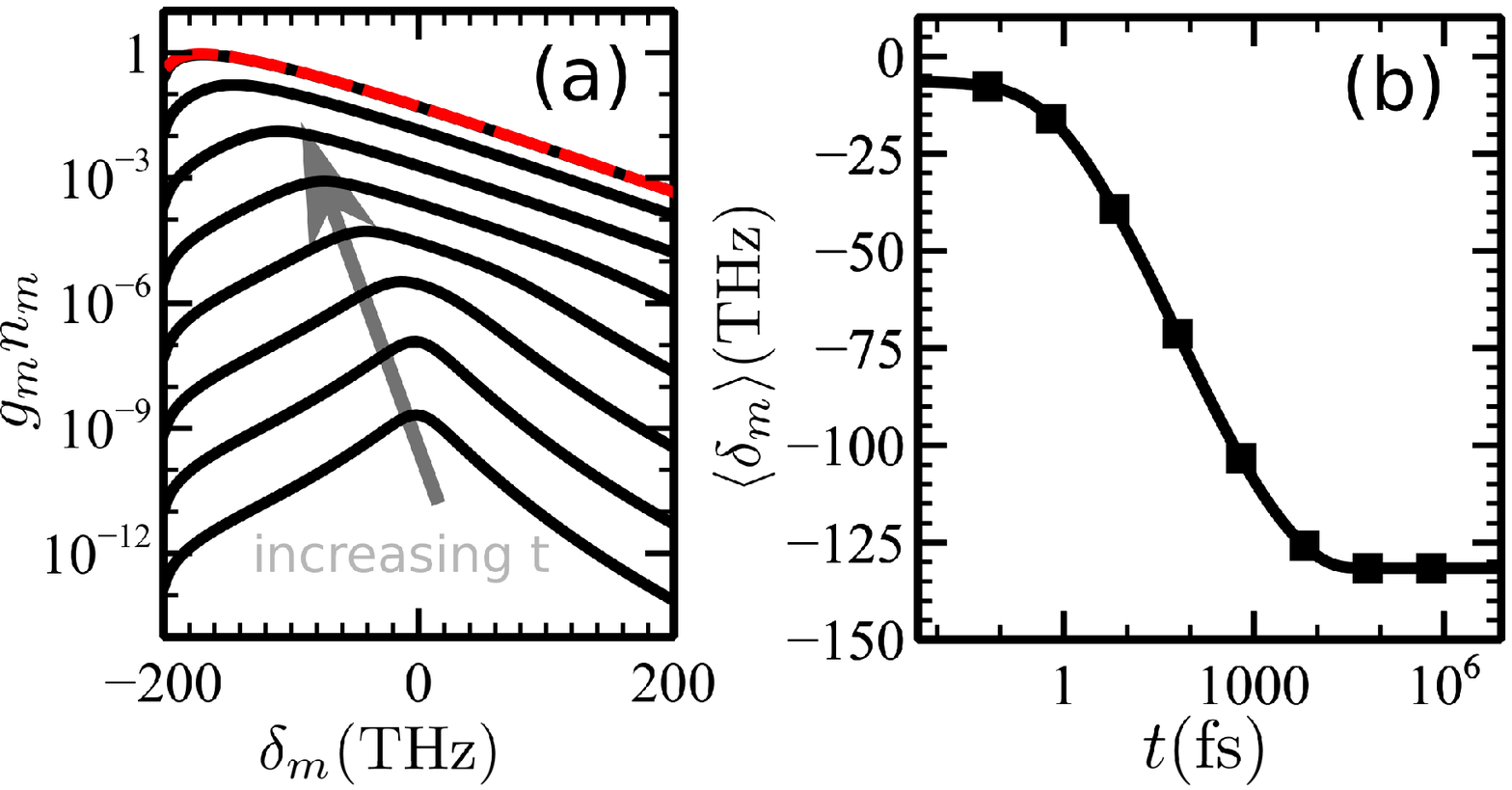}
  \includegraphics[width=3.1 in]{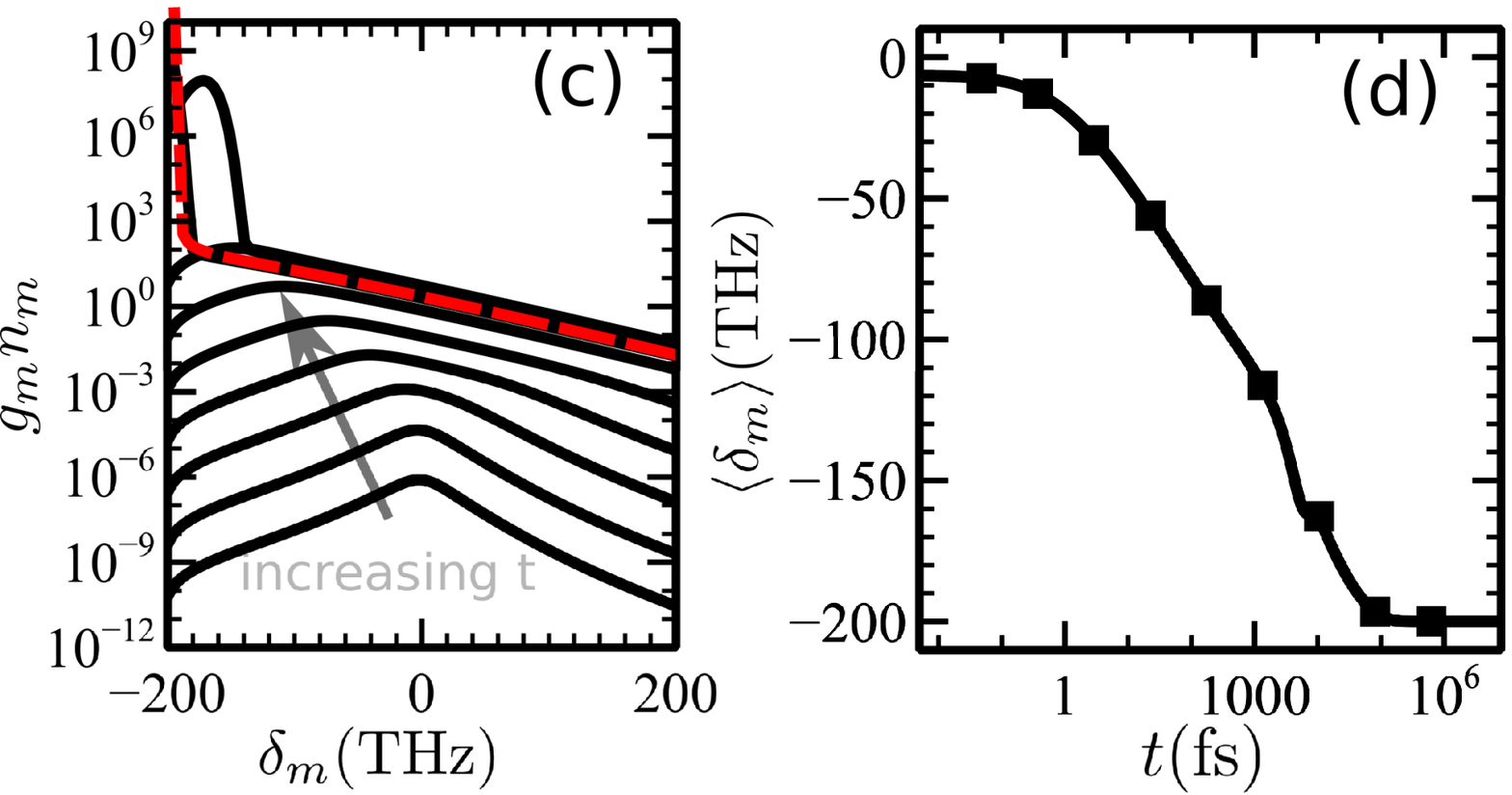}
  \caption{(Color online) Time evolution towards thermal equilibrium
    both below ($\Gu=0.2\Gamma_{\rm thresh}$) (a)-(b) and above ($\Gu=10\Gamma_{\rm thresh}$) (c)-(d) threshold. In (a) and (c) we
    show the photon distribution at different times, the dashed red
    curves show the steady state solution. In (b) and (d) we show the
    mean frequency of the photon intensity, $\average{\delta_m}$, as a
    function of time. The squares show the locations at which the
    traces in (a) and (c) are taken. Parameters are the same as in Fig.\ \ref{fig:Ntotinvmu}.}\label{fig:timeev}
\end{figure}

By integrating Eqs.~\eqref{eqn:nmdot}-\eqref{eqn:pedot} we can examine
the time evolution of the system towards the steady state thermal
equilibrium distribution. Examples of this are shown in
Fig.~\ref{fig:timeev}, where we evolve the equations in time by
turning on the pump at $t=0$. We may then follow the evolution from an
initially empty cavity to the final Bose-Einstein distribution. In
Fig.~\ref{fig:timeev}(a) and (c) we plot the photon distribution at
various times as solid curves along with the equilibrium distribution
as a dashed red curve.  In Fig.~\ref{fig:timeev}(b) and (d) we
plot the average energy of the photon distribution,
$\average{\delta_m}$ as a function of time.  The markers on these
curves indicate the times at which the traces in (a) and (c) were
taken.

The first light emitted into the cavity modes follows closely the
  bare fluorescence spectrum, and so is dominated by photons with
  energies close to $\delta_m=0$.  Thus, at these early times no
  thermalization is seen, and the same profile appears both above and
  below threshold.  It is important to note that these early
  times are already longer than the timescale $1/\gamma$
  required for the rovibrational states to reach thermal
  equilibrium. For the parameters used in Fig.~\ref{fig:timeev} this
  rovibrational thermalization timescale is only $0.1$fs.  The photons
  take longer to reflect this thermal distribution because
  thermalization of photons requires the balance of emission and
  absorption processes; i.e.\ it is only as photons are absorbed and
  re-emitted that a thermal distribution emerges from the balance
  between these processes.  As this occurs, the mean of the
distribution starts to shift towards the low energy modes of the
cavity. Above threshold, as the equilibrium distribution is reached,
the thermal tail slightly overshoots its steady state value and a
large population appears in a mode slightly above $\delta_0$ before
this finally moves to the ground mode. The thermalization time for
each mode is set by $\Gamma(\delta_m)$ and so is longer for the modes
furthest away from the molecular transition frequency. The onset of a
macroscopic occupation in the ground mode is accompanied by a kink in
the mean photon frequency which occurs at the point where the
macroscopic peak first appears. Finally, in both cases, the mean
settles to its stationary value (very close to $\delta_0$ above
threshold) as the steady state is reached. The themalization time
which we find for these experimentally realistic parameters is of
order 10-100ps, which is similar to the measured
result~\cite{Schmitt2014b}. 

{\section{Beyond the Rate Equation Model} \label{sec:linewidths}

  To look at the correlations in the system it is necessary to go
  beyond the rate equation description above. The rate equations
  neglect the correlations between the light and dye beyond
  first order, and so are unable to capture the behavior of the
  variances and higher order moments of the distribution. In this
  section we will show how, using a master equation for the full
  probability distribution, one may calculate both the second
  order photon coherence, $g^{(2)}$, and the linewidth of the
  emission.

  To simplify the calculation it is instructive to look at the single
  mode version of the model considered so far. To do this we ignore the
  modes which make up the thermal tail and concentrate only on the
  mode which condenses.  In this case the master equation is given by
\begin{multline}
  \label{eqn:MEsingle}
  \dot{\hat\rho} =-i\delta[ \ad \hat{a},\hat\rho] -
    \frac{\kappa}{2}\mathcal{L}[\hat{a}]\hat\rho-\sum_{i}\left\{\frac{\Gu}{2}\mathcal{L}[\hat{\sigma}_i^+]+\frac{\Gd}{2}\mathcal{L}[\hat{\sigma}_i^-]\right.\\
    \left.+\frac{\Gamma(-\delta)}{2}\mathcal{L}[\ad\hat{\sigma}^-_i]+\frac{\Gamma(\delta)}{2}\mathcal{L}[\hat{a}\hat{\sigma}^+_i]\right\}\hat\rho.
\end{multline} 
To proceed we note that the steady state of the above equation can
  be written in the form $\hat\rho = \sum_{n,m} P_{n,m} \ket{n}\!\bra{n} \otimes \hat\rho_m$ where $\ket{n}$ is a photon number state and
  $\hat\rho_m = \sum_{\{ s_i =0,1\}} \delta_{m, \sum_i \! s_i}
  \bigotimes_i |s_i\rangle \langle s_i|$ is the incoherent mixture of
  all molecular states with a total number of excited molecules $m$.
  This steady state is thus ``diagonal'' in the number space of
  photons and excited molecules, and the quantity $P_{n,m}$ gives the
  probability of finding $n$ photons and $m$ electronic excitations in
  the system.  One can easily check that such an ansatz exactly
  satisfies Eq.~(\ref{eqn:MEsingle}) as long as $P_{n,m}$
  obeys the equation:
\begin{multline} \label{eqn:Ptreal}
  \dot{P}_{n,m} = \kappa [(n+1)P_{n+1,m} - nP_{n,m}] \\
  \begin{aligned}
  +& \Gamma_{\uparrow}[(N-m+1)P_{n,m-1}-(N-m)P_{n,m}] \\
  +& \Gamma_{\downarrow}[(m+1)P_{n,m+1}-mP_{n,m}] \\
  +& \Gamma(-\delta)[n(m+1)P_{n-1,m+1} -(n+1)mP_{n,m}] 
  \end{aligned} \\
  + \Gamma(\delta)[(n+1)(N-m+1)P_{n+1,m-1} - n(N-m)P_{n,m}].
\end{multline}
Such a form is sufficient because the full quantum master equation
  in Eq.\ \eqref{eqn:ME} is written within a secular approximation.
  This means that the evolution of diagonal and off-diagonal terms can
  be separated, as has been done here.

\subsection{Second order coherence $g^{(2)}$}

We can the use this equation for $P_{n,m}$ to calculate the zero
time delay second order quantum coherence of the emitted light field
$g^{(2)}(0)$.  This quantity is defined as
\begin{equation}
  g^{(2)}(0)=1+\frac{\sigma_n^2-\average{n}}{\average{n}^2},
\end{equation}
where we define averages and variances as
$\average{X}=\sum_{n,m}XP_{n,m}$ and
$\sigma_X^2=\average{(X-\average{X})^2}$ respectively.  Such a
correlation function has been experimentally
measured~\cite{Schmitt2014} where it was found to smoothly cross over
from thermal light with $g^{(2)}=2$ below threshold to coherent light
with $g^{(2)}=1$ far above threshold. Below we show that such
  behavior is indeed reproduced by Eq.~(\ref{eqn:Ptreal})

  For small system sizes we can solve Eq.  \eqref{eqn:Ptreal}
  numerically to find the steady state probability distribution.  This
  then allows one to extract the required moments to find
  $g^{(2)}$. Unfortunately, direct numerical solution of
  Eq.~(\ref{eqn:Ptreal}) is only tractable for relatively small
  numbers of molecules $N$.  The value of $N$ sets the variances of
  both $n$ and $m$, and so with increasing $N$, the number of non-zero
  values of $P_{n,m}$ grows.  This means that brute force numerics is
  not feasible with realistic values of $N \simeq 10^9$.  In this
  large system size limit we can however get a good approximation to
  the behavior of the system above threshold by writing expressions
  for the moments and truncating at second order. This is equivalent
  to assuming that the full probability distribution is Gaussian.
  Such an assumption is reasonable  above threshold,
  but fails at low pump powers.  For a Gaussian distribution, one need
  only calculate the first and second moments of the distribution, as
  all  moments factorize.  The equations of motion for the first
  moments, $\partial_t \langle n \rangle, \partial_t \langle m
  \rangle$ give:
\begin{align}
  0 =& 
  - \kappa \average{n} 
  + \Gamma(-\delta) [(\average n+1) \average m + \sigma^2_{nm}]
  \nonumber\\ 
  &- \Gamma(\delta) [\average n (N-\average m) - \sigma^2_{nm}],
  \\
  0 =& \Gamma_{\ua}(N-\average m) - \Gamma_{\da} \average m
  \nonumber\\ 
  &- \Gamma(-\delta) [(\average n+1) \average m + \sigma^2_{nm}]
  \nonumber\\ 
  &+ \Gamma(\delta) [\average n (N-\average m) - \sigma^2_{nm}],
\end{align} while the equations of motion for the second order
moments give rise to the following conditions
\begin{widetext}
\begin{gather}
\begin{split}
  0 = 
  - \kappa \left( - \average n +2 \sigma_n^2 \right)
  + \Gamma(-\delta) \left[ (\average n+1) \average m
    + 2 \sigma^2_n \average m + \sigma^2_{nm} (2 \average n + 3) \right]
  \\- \Gamma(\delta) \left[-\average n(N-\average m)
    + 2 \sigma_n^2 (N-\average m) - \sigma^2_{nm}(2\average n-1) \right],
\end{split}
    \\
    \begin{split}
  0 = \Gamma_\ua \left[  (N-\average m) - 2 \sigma_m^2 \right]
  - \Gamma_\da \left[- \average m + 2 \sigma_m^2 \right]
  - \Gamma(-\delta) \left[ - (\average n+1)\average m
    + 2 \sigma_m^2 (\average n+1) + \sigma^2_{nm} (2 \average m -1) \right]
  \\+ \Gamma(\delta)\left[ \average n  (N-\average m)
    - 2\sigma^2_m \average n + \sigma^2_{nm} (-2 \average m + 2N-1) \right],
    \end{split}
\\
\begin{split}
  0 = - (\kappa+\Gamma_\da+\Gamma_\ua) \sigma^2_{nm}
  + \Gamma(-\delta) \left[ - (\average n+1) \average m 
    + \sigma^2_m (\average n+1) - \sigma_n^2 \average m
    + \sigma^2_{nm}(\average m -  \average n -2) \right]
 \\ + \Gamma(\delta) \left[ - \average n (N-\average m) 
    + \sigma^2_m \average n + \sigma^2_n (N- \average m)
    + \sigma^2_{nm}( \average m - \average n + 1 -N) \right].
    \end{split}
\end{gather}
\end{widetext}
Here we have introduced the covariance of the distribution
$\sigma^2_{nm}=\average{(n-\average{n})(m-\average{m})}$. It is
 straightforward to numerically solve these
equations, regardless of the value of $N$.  Indeed, for large $N$
  the expressions can be further simplified by making an expansion in
  $1/N$, noting that both first moments and variances all scale
  linearly with $N$.

\begin{figure}
  \centering
  \includegraphics[width=2 in]{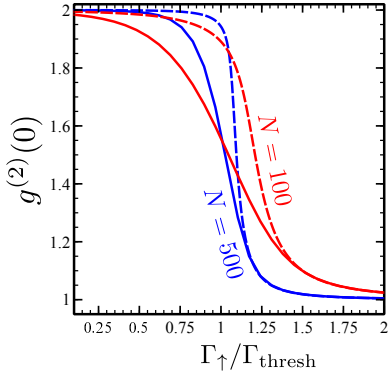}
  \caption{(Color online) Second order coherence function $g^{(2)}$ as we sweep through threshold. The solid red curve shows the results of a full numerical calculation of the probability distribution for $N=100$ while the red dashed curve shows the result of the calculation based on second order cumulants described in the text. The blue curves show similar results for $N=500$. The other parameters used are the same as those in Fig.~\ref{fig:Ntotinvmu} except we directly specify $\Gamma(-\delta)=1$GHz, $\Gamma(\delta)=50$MHz.}\label{fig:g2}
\end{figure}

In Figure~\ref{fig:g2} we plot the value of $g^{(2)}$ calculated by
solving the full master equation for both $N=100$ and $N=500$ and
compare these with the results of the second moment calculation
described above. For both calculations we see a crossover from $g^{(2)}=2$ to $g^{(2)}=1$ which becomes sharper as the number of molecules is increased. This is the same behaviour as is observed experimentally~\cite{Schmitt2012a}.The two approaches
match well above and at threshold, however, at small pumping
strengths the crossover is not captured correctly by the second order cumulants. This is because below threshold the probability distribution
of the system is far from Gaussian and so the approach based on
truncating at second order breaks down.

As the number of
  molecules grows, the transition between $g^{(2)}=1$ and
  $g^{(2)}=2$ becomes increasingly sharp; this is seen by both the
  full probability distribution and the cumulant calculation.  For
  $N=10^9$, the cumulant calculation predicts a very sharp transition
  at threshold.  In comparing these results to the recent
  experiments~\cite{Schmitt2014} it should be noted that in this
  section we have considered a single-mode approximation; calculating
  the full non-equilibrium correlation function for a many mode system
  is a challenge for future work.

\subsection{Emission lineshape}

To examine the temporal coherence of the emitted light source we can use a similar formalism to calculate the emission spectrum of the cavity~\cite{walls:94}
\begin{equation} \label{eqn:adaspec}
	S_{\ad \hat{a}}(\omega)=2 {\rm Re}\int_{0}^\infty dt \average{\ad(t)\hat{a}(0)}{\rm e}^{i\omega t}.
\end{equation}
From the quantum regression theorem~\cite{walls:94} we know that the
equation of motion for this two-time correlation function is simply
given by the evolution of $\ad(t)$ starting from the initial density
matrix $\hat\rho(0)=\hat{a}\hat\rho_{ss}$ where
$\hat\rho_{ss}$ is the stationary state. this means that we need to
find the time evolution of
\begin{equation}
  \average{\ad(t)\hat a(0)}=\sum_{n,m}\sqrt{n}P^1_{n,m}, \label{eqn:adaPn1}
\end{equation}
where $P^1_{n,m}$ represents the elements of the density matrix
which are on the first off-diagonal in the photon number basis,
  and diagonal in number of excited molecules.  These correspond to
  defining ${\hat\rho} = \sum_{n,m} P^1_{n,m} \ket{n-1}\bra{n}
  \otimes \rho_m$ in an analogous way to the definition of $P_{n,m}$
  in the previous section. The equation for the evolution of this can
be derived from the master equation in the same way as Eq.\ \eqref{eqn:Ptreal} and is given by
\begin{multline} \label{eqn:P1}
	\dot{P}^1_{n,m} = \kappa [\sqrt{n(n+1)}P^1_{n+1,m} - (n-1/2)P_{n,m}^1] \\
	+ \Gu[(N-m+1)P^1_{n,m-1}-(N-m)P^1_{n,m}] \\
	+ \Gd[(m+1)P^1_{n,m+1}-mP^1_{n,m}] \\
	+ \Gamma(-\delta)\left[\sqrt{n(n+1)}(m+1)P^1_{n-1,m+1} -(n+1/2)mP^1_{n,m}\right] \\
	+ \Gamma(\delta)\left[\sqrt{n(n+1)}(N-m+1)P^1_{n+1,m-1}\right. \\ \left. \vphantom{\sqrt{n(n+1)}}- (n-1/2)(N-m)P^1_{n,m}\right].
\end{multline}
Here we have ignored the Hamiltonian terms since these only shifts the origin of frequency in the power spectrum. The full problem then reduces to finding the time evolution of the above equation using the initial condition $P^1_{n,m}(0)=\sqrt{n}P_{n,m}^{\text{steady}}$. The same numerical techniques as before can be applied to this problem to find the emission spectrum. 

In Figure \ref{fig:linewidths} (a) and (b) we show the emission
spectrum below and above threshold respectively, calculated
numerically for $N=100$, along with Lorentzian fits to the data. We
see that above threshold the lineshape is almost perfectly Lorentzian
while below threshold there is some deviation at high frequencies,
this is because below threshold there is a significant effect from the
dynamics of the molecules on short timescales which causes the
lineshape to deviate slightly.  Above threshold, there is a
significant separation of timescales, so that the slow
exponential decay of correlations occurs on much longer timescales
than the fast dynamics of the molecular state, and so the two can
be clearly separated --- note the frequency scales appearing in
Fig.~\ref{fig:linewidths}.

\begin{figure}
  \centering
  \includegraphics[width=3.2 in]{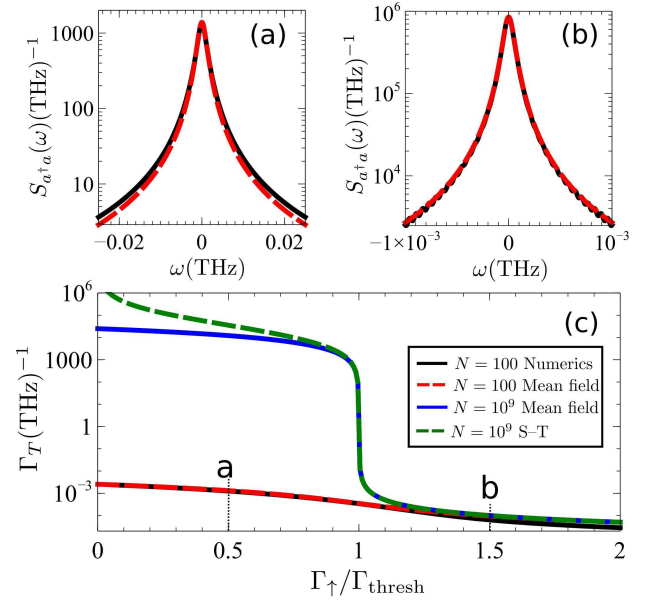}
  \caption{(Color online) The emission spectra below (a) and  above (b) the transition point, calculated for $N=100$. The exact solution is shown as the solid (black) lines while the Lorentzian fits are shown as dashed (red) curves. Panel (c) shows the linewidth of the emission spectrum. In the small system size limit ($N=100$) we show the full numerics as the solid black line while the mean field approximation is given by the red dashed line. In the thermodynamic limit ($N=10^9$) we show the full mean field calculation as the blue solid line while the Schawlow--Townes result is the green dashed curve. The points marked a and b indicate the locations at which the spectra in the other panels were calculated. The other parameters used are the same as those in Fig.~\ref{fig:Ntotinvmu} except we directly specify $\Gamma(-\delta)=1$GHz, $\Gamma(\delta)=50$MHz.}\label{fig:linewidths}
\end{figure}

As above, for large $N\simeq 10^9$, exact solution of
  Eq.~(\ref{eqn:P1}) is no longer feasible.  To find the
  linewidth in the thermodynamic limit we can however make a
mean field approximation. This can be done by calculating an explicit
expression for $\average{\ad(t)\hat a(0)}$ from Eq.\
\eqref{eqn:adaPn1} and truncating the resulting evolution at the mean
field level, i.e.  assuming that $P^1_{n,m} \simeq P^1_n P_m$ can
  be factorized.  We find for the correlation function
\begin{equation} \label{eqn:ada}
	\frac{d}{dt}\average{\ad(t)\hat{a}(0)}=-\left(\frac{\Gamma^-(t)}{2}-\frac{\Gamma^+(t)}{2}\right)\average{\ad(t)\hat{a}(0)}.
\end{equation}
where the rates are given by
\begin{gather}
	\Gamma^-(t)=\kappa+\Gamma(\delta)(N-\average{m}), \\
	\Gamma^+(t)=\Gamma(-\delta)\average{m}.
\end{gather}
Here the coupling between the photonic and molecular degrees of
freedom means that the rates which occur in this equation depend on
the inversion of the molecules. This dependence encodes
the short-time effects discussed above, whereby the state of
the molecules is perturbed by the removal of one photon.
It is therefore necessary to follow the evolution
of $\average{m(t)}$ from its initial steady state value,
according to its equation of motion,
\begin{gather}
	\dot{\average{m}}=\tGu(t) (N-\average{m})-\tGd(t) \average{m}.
\end{gather}
The rates in this expression depend on the current photon population:
\begin{gather}
	\tGd=\Gd+\Gamma(-\delta)(\average{n}+1), \\
	\tGu=\Gu+\Gamma(\delta)\average{n},
\end{gather}
and $\average{n}$ evolves according to
\begin{gather}
	\dot{\average{n}}=-\Gamma^-(t)\average{n}+\Gamma^+(t)(\average{n}+1).
\end{gather}
This then gives a closed set of equations for the evolution which can be used with the quantum regression theorem to calculate the correlation function.

As already noted earlier, above threshold there is a separation of
  timescales between the fast molecular dynamics and the slow decay of
  correlations.  This means that for the purpose of finding an
  approximate expression for the lineshape above threshold, we may
  ignore the (fast) time dependence of the rates in
Eq.~\eqref{eqn:ada} and simply using the steady state values. This
gives rise to the Lorentzian spectrum
\begin{equation} \label{eqn:lorentzian}
	S_{\ad \hat{a}}(\omega)=\frac{2\average{n}_{ss}\Gamma_T}{(\delta+\omega)^2+\Gamma_T^2},
\end{equation}
where $\Gamma_T=[\Gamma^-(\infty)-\Gamma^+(\infty)]/2$ is the
effective decay rate of the two-time correlation function evaluated in
the steady state.

We can then use this calculation to look at the linewidth of the
emission spectrum calculated using Eq.~\eqref{eqn:adaspec}. An example
of the way in which the width of the emission spectrum changes as the
pump power is increased through the transition threshold is shown in
Fig.~\ref{fig:linewidths} (c). For small particle numbers the full
numerics agree very closely with the results of the mean field
calculation with only a slight deviation above threshold where the
variances become important. This then allows us to trust the results
of the mean field calculation in the thermodynamic limit. For $N=10^9$
we see that below threshold the emission is very broad and weak. At
the transition point the linewidth collapses and then saturates at a
value which is controlled both by the losses from the cavity and the
absolute value of the dye absorption and emission rates. The below
threshold linewidth is controlled by the number of molecular
exitations, in this limit there is no stimulated emission and the
emission is broadened by the molecules.

Despite the excellent match to the linewidth,
it is worth noting that the mean field model presented here cannot
accurately calculate the full spectrum above threshold. In this
limit the dynamics goes through a region where there are correlations
between the photons and the molecules and hence the factorization of
the cumulants is no longer valid, in this region the mean field model
predicts unphysical negative spectral weight at some
frequencies.  Such considerations only effect the short time
dynamics, and not the slow dynamics that determine the linewidth.
The weight of the Lorentzian lineshape is however modified by
this early time dynamics.

In the large photon number limit the linewidth given by $\Gamma_T$
  ultimately takes the
  Schawlow--Townes~\cite{Schawlow1958,Scully1997} form:
\begin{equation}
	\Gamma_T=\frac{\kappa}{2}+\frac{N[\Gamma(\delta)\Gamma(-\delta)-\Gu\Gamma(-\delta)+\Gd\Gamma(\delta)]}{2n_{ss}[\Gamma(\delta)+\Gamma(-\delta)]}.
\end{equation}
This is plotted alongside the mean field result in
Fig.~\ref{fig:linewidths} as the dashed (green) line, we see
good agreement in the condensed phase which breaks down, as expected,
when the occupation of the photon mode is small.

\section{Conclusions} \label{sec:conc}

In this paper we have developed a quantum mechanical model capable of
describing the thermalization of photons inside a dye-filled
microcavity. We have shown how this model is able to predict the behavior of the
recent experiments on photon Bose-Einstein condensation. 

From our full quantum
model we derived a rate equation capable of describing
many features of the system. We were able to define a threshold
condition which allows us to identify when the system transitions to a
macroscopically occupied state. This was then used to investigate the
breakdown of thermalization in the photon condensate. We showed how,
by changing the length of the cavity, the low energy modes interact
with the dye too weakly to thermalize and the Bose-Einstein description
breaks down. We also looked at how, for extreme parameters of the dye,
it is possible to make the system selectively thermalize only in
certain frequency regimes. This led us to identify two possible
criteria which can identify if the system is close to thermal
equilibrium or not: the mode which gains a macroscopic occupation and
the chemical potential of a Bose-Einstein distribution fit to the
above threshold distribution. The phase diagrams of these criteria
  for thermalization as a function of temperature and parameters of
the dye were then examined. We have investigated the way in which
the system approaches a thermal equilibrium distribution by examining
the dynamics as the pump is switched on. 

We have also begun to explore the quantum correlations of such a
system, going beyond the mean-field (i.e.\ rate equation)
  description. Using the full quantum model we have shown how one may
calculate both the second order coherence of the emitted light and its
  linewidth.

\acknowledgments

{We are very glad to acknowledge stimulating discussions with
  M.\ Weitz, J.\ Klaers and R.\ Nyman.  The authors acknowledge
financial support from EPSRC program ``TOPNES'' (EP/I031014/1) and
EPSRC (EP/G004714/2). PGK acknowledges support from EPSRC (EP/M010910/1).

\bibliographystyle{apsrev4-1}
\bibliography{PhotonCondensates}

\end{document}